\documentclass[%
 reprint,
 amsmath,amssymb,
 aps, nofootinbib
]{revtex4-2}

\usepackage{macros} 
\usepackage{caption}
\usepackage{subcaption}
\usepackage{graphicx} 
\usepackage{dcolumn}  
\usepackage{xcolor}   
\usepackage{bm}       
\usepackage{multirow} 
\usepackage{array}

\definecolor{blue}{rgb}{0.36, 0.54, 0.85}
\definecolor{amaranth}{rgb}{0.9, 0.17, 0.31}
\definecolor{pink}{rgb}{0.87, 0.56, 0.81}
\definecolor{ao}{rgb}{0.0, 0.5, 0.0}
\definecolor{maroon}{rgb}{0.76, 0.13, 0.28}
\definecolor{cardinal}{rgb}{0.77, 0.12, 0.23}
\definecolor{lightcardinal}{rgb}{0.97, 0.42, 0.53}
\definecolor{frenchlila}{rgb}{0.53, 0.38, 0.56}
\definecolor{yellow}{rgb}{1.0, 1.0, 0.87}
\definecolor{lightseagreen}{rgb}{0.45, 0.85, 0.58}
\definecolor{gray}{rgb}{0.9, 0.9, 0.9}
\definecolor{lightblue}{rgb}{0.66, 0.84, 0.96}

\usepackage[pagebackref=false,
			hyperindex=true,
			citecolor=blue,
			colorlinks=true,
			linkcolor=blue,
			urlcolor=blue]{hyperref}

\usepackage{hyperref}

\newcommand{\param}{\ensuremath{\boldsymbol{\theta}}}

\newcommand\abs[1]{\left|#1\right|}

\newcommand{\mrm}[1]{\mathrm{#1}}

\begin{document}


\title{Heavy-tailed likelihoods for robustness against data outliers: \\ 
Applications to the analysis of gravitational wave data}

\thanks{\href{mailto:asasli@auth.gr}{asasli@auth.gr} }

\author{Argyro Sasli$^{1,\ast}$, Nikolaos Karnesis$^{1}$, Nikolaos Stergioulas$^{1}$\\
$^{1}$ Department of Physics, Aristotle University of Thessaloniki, Thessaloniki 54124, Greece}%

\date{\today}

\begin{abstract}
In recent years, the field of Gravitational Wave Astronomy has flourished. With the advent of more sophisticated ground-based detectors and space-based observatories, it is anticipated that Gravitational Wave events will be detected at a much higher rate in the near future. One of the future data analysis challenges is performing robust statistical inference in the presence of detector noise transients or non-stationarities, as well as in the presence of stochastic Gravitational Wave signals of possible astrophysical {\color{black}and/or cosmological} origin. The incomplete knowledge of the total noise of the observatory can {\color{black}introduce challenges} in parameter estimation of detected sources. In this work, we propose a heavy-tailed, {\color{black} Hyperbolic likelihood}, based on the Generalized Hyperbolic distribution. With the Hyperbolic likelihood we obtain a robust data analysis framework against data outliers, noise non-stationarities, and possible inaccurate modeling of the noise power spectral density. We apply this methodology to examples drawn from gravitational wave astronomy, and in particular to synthetic data sets from the planned LISA mission.
\end{abstract}

\maketitle


\section{Introduction\label{sec:introduction}} 

In recent years and since the first detection of a Gravitational Wave (GW) signal~\cite{firstGW}, more than 90 GW signals from the inspiral and merger of compact astrophysical objects have been included in published catalogues of GW detections achieved by the LIGO~\cite{LIGO} and Virgo~\cite{Virgo} ground-based detectors~\citep{GWTC,GWTC1,GWTC-2,GWTC-4,GWTC-3,GWTC-5, Nitz}. In the near future, the existing detectors (including KAGRA~\cite{KAGRA}) will operate at improved sensitivity, while the construction of LIGOIndia~\cite{Unnikrishnan, Saleem_2022} is expected to start. In the meantime, a new generation of detectors (Einstein Telescope (ET)~\cite{ET}, Cosmic Explorer (CE)~\cite{CE}  and NEMO~\cite{NEMO}) is in the planning phase, and the space-borne Laser Interferometer Space Antenna (LISA)~\cite{Amaro2017} and TianQin and Taiji~\cite{chinesedetectors} are expected to operate after the mid-2030s. Each one of these observatories is based on different designs, which translates to different sensitivity curves aiming at different frequency ranges for the detected GW signals. One of their similarities, however, is that all 3rd-generation detectors will be characterized by higher rates of GW signals, with a significant fraction overlapping in time and frequency~\cite{Wu_Nitz}. At the same time, instrumental noise knowledge is crucial to successfully detect and characterize GW signals~\cite{PhysRevD.107.063004, RevModPhys.94.025001, Littenberg:2014oda, Baghi2023qnq, Speri:2022kaq}. Detector noise non-stationarities, such as slow noise Power Spectral Density (PSD) variations~\cite{PhysRevD.102.084062}, instrumental transients (glitches)~\cite{LISAPathfinder:2022awx, Baghi:2021tfd}, data gaps~\cite{Dey:2021dem,Baghi:2019eqo}, or noise bursts~\cite{PhysRevD.67.082003}, may lead to biased results or even to wrong false alarm rates. Ultimately, we expect noise non-stationarities to play a more important role in analyzing future science-rich data sets. 

In this work, we focus on the case of the LISA mission and the~$\mrm{mHz}$ range of the GW spectrum. We have made this choice because it is certain that resolving the LISA noise budget will be a challenging task~(e.g., see \cite{PhysRevD.107.063004, karnesis21} and references therein). In addition, for the case of ground-based detectors, dedicated experiments can be performed on-site to calibrate the relevant noise models with high accuracy~\cite{Abbott_2020_det}. This practice is inaccessible in space.

LISA is currently in development and is expected to start collecting data after the mid-2020s. This means we currently do not have sufficient prior information {\color{black}about} the possible outliers of the LISA noise. However, we can extrapolate from the LISA Pathfinder (LPF) {\color{black}mission data~\cite{Armano:2018kix, Armano:2016bkm, 2018PhRvD..98j2005A}}. {{\color{black}During the LPF mission, the differential acceleration noise level between $1$ and $100$~$\mrm{mHz}$ was measured to be constantly reducing with time} due to the constantly varying vacuum conditions inside the test-masses caging, throughout the duration of the mission~\cite{Armano:2018ucz, Armano:2018kix}. {{\color{black} Such an effect is also expected in the LISA noise data. Another example of a non-stationary stochastic GW signal in the LISA band will be generated by the ensemble signal} of the Ultra Compact Binaries (UCBs) in the vicinity of our Galaxy~\cite{Amaro2017,PhysRevD.107.063004,karnesis21,georgousi22,korol22}. This confusion signal will have cyclo-stationary properties due to the orbit of the LISA constellation facing in and out of the Galactic center, where most of these objects are located. More details for this type of signal are mentioned in Section~\ref{sec:gws}. Besides the UCBs, we also expect to detect stochastic signals originating from other types of astrophysical sources~\cite{Babak:2023lro, Pozzoli:2023kxy}.

Another noise source that appears as a non-stationary feature in the overall PSD of the noise is the so-called tilt-to-length coupling effect (TTL). {\color{black}The TTL was first measured during the LPF operations}, and it corresponds to the effect of space-craft jitter motion picked up by the sensitive interferometer measurement~\cite{Armano:2016bkm}, generating a ``bump''-like noise between $10-100~\mrm{mHz}$~\cite{Armano:2016bkm, Armano:2018ucz} {\color{black}in the differential acceleration spectrum}. {\color{black}This effect} depends on the geometrical layouts of the optical components of the instruments~\cite{Hartig:2022htm, 2023arXiv230802398A}, and can vary with time depending on environmental {\color{black}factors}~\cite{Armano:2016bkm}. {\color{black}A TTL noise contribution} is predicted for the LISA mission as well. {\color{black}Currently, the characterization plan is based on partially subtracting the effect of TTL} in the post-processing of the data~\cite{Paczkowski:2022nrt, George:2022pky, Houba:2022vwq}. Finally, the experience with LPF data  also provided us with some insight into noisy glitches~\cite{LISAPathfinder:2022awx, Baghi:2021tfd}. 

So far, several techniques have been developed to address each of the issues caused by noise non-stationarities. For example, one way to tackle the problems induced by rapid noisy transients is to model them and fit them simultaneously with the waveform model from the data~\cite{Cornish:2014kda}. This type of analysis is based on trans-dimensional methods, such as the Reversible-Jump Markov Chain Monte Carlo (MCMC)~\citep{RJMCMC, eryn}. {\normalfont Following this methodology, a waveform model is being searched in the data together with the unknown number of models representing the glitches. Then, the parameter space becomes dynamic since the number of such events is unknown.} {\normalfont In the context of LISA data analysis, there have been quite a few applications of this technique~\cite{PRD_RJMCMC_1,PRD_RJMCMC_2,PRD_RJMCMC_3, Baghi_2023}}. Another approach was introduced in~\cite{Edwards:2020tlp}, which focused on the idea of testing the stationarity of the time series. Based on a surrogate data approach, this framework begins with detecting the data segments with different statistical noise properties. Then, a flexible PSD model is used  to fit the different sections of the data~\cite{PhysRevD.92.064011}.

Finally, there is the approach of tackling this kind of problem  {\color{black}at the level of the adopted likelihood function, where models with wider tail properties than the usual Gaussian distribution can be used}. Such a solution was proposed with Student's t-distribution~\cite{roever2011A, roever2011B, Banagiri_2020}, {\color{black}which has been} applied to GW data by the LVK collaboration. In~\cite{roever2011A}, one begins by adopting the $\mrm{Inv}-\chi^2$ prior distribution for each of the PSD coefficients of the noise, which is conjugate with respect to the Gaussian likelihood. Then, the resulting joint conditional density takes the form of Student's t-distribution with $\nu > 0$ degrees of freedom, which, depending on the tuning parameter $\nu$, has the desired heavier tails compared to the Gaussian distribution. This feature can accommodate any deviations of the data with respect to our modeled PSD of the noise, whether these originate from fast noisy transients or other types of noise non-stationarities. In addition, in~\cite{Banagiri_2020}, it was found that the uncertainties in the phase evolution of the binary neutron star signal could be accommodated in addition to the uncertainty in the noise. {\normalfont However, following this approach, one should take into account that the choice of prior can have a non-negligible impact on the final posterior.}

Besides the above examples, one can also follow the recipe of~\cite{Martellini:2014xia}, where the non-Gaussianity of the data is modeled  using a higher-order Edgeworth expansion for the unknown signal distribution. Other works~\cite{Hamimeche:2008ai,Verde_2003, Flauger:2020qyi}, have adopted a linear combination of the Gaussian and the Log-Normal likelihoods. This was mostly done in order to accommodate  the systematic bias due to data averaging. Older works also proposed using different linear combinations of distributions for the likelihood function~\cite{PhysRevD.60.021101, PhysRevD.65.122002}.

{\color{black}Here}, we introduce {\color{black} the Hyperbolic likelihood $\Lambda_{\cal H}$, which is based on a special case of the} Generalized Hyperbolic ($\cal GH$) distribution, for the analysis of GW data. The $\cal GH$ distribution has been used in the statistical description of relatively short time series in the field of Finance {\color{black}and Econometrics}~\cite{Bianchi2020, Borak2011,Prause1997ModellingFD, Eberlein2002TheGH, Prause1999TheGH, KUCHLER19991}. The $\cal GH$ distribution has the advantage of being able to arrive to many known distributions from the exponential family~\cite{Prause1999TheGH}, by simply tuning its overall shape to adjust to the given {\color{black}statistical properties of the input} data. Thus, for example, while recovering the parameters of a transient signal, we can simultaneously recover the statistical properties of the underlying noise. 

{\normalfont In practice, we increase the dimensionality of the problem by a couple of parameters, which control the shape of the distribution. Those are estimated from the residual data, and are then used to infer their actual distribution, simply because the Hyperbolic distribution has a greater flexibility and can adjust to different distributions with perfect or acceptable accuracy, as we demonstrate. For example, if the residual data have a Gaussian distribution, then the parameters of the Hyperbolic distribution will converge to values that agree with a Gaussian distribution asymptotically (see section~\ref{sec:generalized_hyperbolic_distr} for details).}

{\normalfont Consequently, the main and most important advantage of this framework is that it comprises a robust approach for both parameter estimation for the transient signals and the characterization of the underlying noise. This feature is crucial in cases where the detector noise is ``polluted'' with noise transients, or even when the noise model is not completely known.} As already mentioned, such will be the case of the signal-dominated LISA data, where the total noise will be the sum of the instrumental noise plus the given unresolvable stochastic GW signal~\cite{Amaro2017, LISACosmologyWorkingGroup:2022jok, karnesis21}.

In Section~\ref{sec:background} we introduce the basic theory behind the $\cal GH$ distribution {\color{black} and propose the Hyperbolic likelihood}, placing it in the context of detecting signals in noisy data. Some basic test cases are  discussed and we apply this framework to real data from the LISA Pathfinder mission~\cite{Armano:2016bkm, Armano:2018kix, Armano:2018kix}. In Section~\ref{sec:gws} we apply this formulation to examples in GW astronomy, using synthetic data. We use different scenarios on our assumptions on the noise knowledge of the LISA data channels and then apply this formulation to the case of a single Verification Binary taken from the LISA Data Challenge~\cite{ldc} catalogues. {\normalfont This application assumes cases where the PSD of the noise is considered either unknown or intentionally mischaracterized. In the case of unknown noise, adopting the Hyperbolic likelihood allows us to recover both the waveform and noise parameters. When the noise is assumed to be at the wrong level, the performance analysis based on the Hyperbolic likelihood is identical to the analysis based on the Whittle likelihood, which we use as a benchmark for our analyses.} Finally, in section~\ref{sec:conclusions} we present our conclusions and discuss our findings. 


\section{Theoretical Background\label{sec:background}}

\subsection{The Gaussian distribution case \label{sec:gaussian_distr}}

We begin by assuming that a measurement $y(t)$ is the sum of a signal $h$ that might depend on a parameter set $\param$ and a noise component as
\begin{equation}
y = h(\param) + n ,
\label{eq:data}
\end{equation}
where we have omitted the dependence on time $t$ for the sake of clarity. Then, assuming Gaussian properties of the noise, the likelihood of the measurement $y$ given a parameter set \param, takes the form of 
\begin{equation}
    p(y|\param) 
    {}= C \times e^{ -\frac{1}{2}\langle y - h(\param) | y - h(\param) \rangle},
\label{eq:gaussian}
\end{equation}
{\color{black}where $\langle \cdot|\cdot \rangle$ denotes the noise-weighted inner product between two real time series. In the general multi-dimensional case of $m$ data channels
\begin{equation} 
	\mathbf{a} (t) =
	\begin{pmatrix}
 	a_1 (t) \\
 	a_2 (t) \\
	\vdots \\
	a_{m} (t)
	\end{pmatrix}, \quad \text{and} \quad
	\mathbf{b} (t) =
	\begin{pmatrix}
 	b_1 (t) \\
 	b_2 (t) \\
	\vdots \\
	b_{m}(t)
	\end{pmatrix},
	\label{eq:multchannels}
\end{equation} 
we write the $\langle \cdot|\cdot \rangle$ in matrix form as
\begin{equation}
	\langle \mathbf{a} | \mathbf{b} \rangle = 4 \, \mathrm{Re}\int\limits_0^\infty \mrm{d}f \left[ \tilde{\mathbf{a}}^\dag(f) \mathbf{S}_n^{-1}(f) \tilde{\mathbf{b}}(f) \right],
	 \label{eq:inprod}
\end{equation}
where $\mathbf{S}_n$ becomes the one-sided cross-spectral matrix of the noise for the given arrays of time series measurements. The tilde $(\,\tilde{ }\,)$ denotes the Fourier transform, and the $ (^\dag)$  represents the conjugate transpose operation. } In the end, we conveniently write the log-likelihood as
\begin{equation}
    \Lambda_\mrm{\cal N}(\param) \propto -\frac{1}{2}\langle y - h(\param) \big| y - h(\param) \rangle.
\label{eq:gaussian_llh}
\end{equation}

{\color{black}In Eq. (\ref{eq:gaussian}), one assumes that the noise is Gaussian and thus one can calculate its PSD accurately. But, this will not be always the case. In case of strong glitches or other non-Gaussian features, the PSD calculation will be inaccurate, as it is based on the hypothesis of a Gaussian distribution. Also, in the case of future signal-dominated detectors, the instrumental noise and hence its PSD will be unknown for large parts of the spectrum.}  

Thus, a model of the spectrum of the noise can be adopted and fitted together with the parameters of the signal. Then, the logarithm of the likelihood can be written as
\begin{eqnarray}
	\Lambda_\mrm{\cal W}(\param) &\propto& -\frac{1}{2}\Biggl( \sum_f  \left[ {\ln}\left(S_n (\param_n) \right)\right] \nonumber\\
 && \hspace{0.9cm} + \ \langle y - h(\param_h) | y - h(\param_h) \rangle \Biggr), 
\label{eq:whittle_llh}
\end{eqnarray}
where  $\param=\param_n\sqcup\param_h$, and therefore $\param_n\subseteq\param$ the parameters of the model of the noise. The above expression constitutes the {\it Whittle likelihood} and is approximate for Gaussian and stationary time series~\cite{whittle, choudhuri, roever2011B}. {\normalfont In~\cite{tang2021}, the authors extend the posterior consistency result of~~\cite{2017arXiv170104846K} to non-Gaussian time series, providing a theoretical justification of posterior consistency under mild assumptions on the time series without having to assume Gaussianity. They suggest that this approach can be applied to non-Gaussian time series and may provide accurate spectral density estimates even if the data do not come from a normal distribution, provided that a large sample size is available.} One can also go a step further and assume the uninformative and improper Jeffreys prior for the noise variance, and marginalize the noise spectrum out of the expression of Eq.~(\ref{eq:whittle_llh})~\cite{roever2011A, vitale14}. 

Shortcomings of the Whittle model have been extensively studied in the literature~(e.g. see \cite{choudhuri, roever2011B,contreras} and references therein), but one very relevant situation in GW astronomy is the measurement of time series with high auto-correlation, which could potentially {\normalfont reduce the efficiency of} the likelihood model of Eq.~(\ref{eq:whittle_llh}). 

{\color{black}As already mentioned in the introduction}, a proposed solution to counterbalance those shortcomings would be to follow the strategy of~\cite{roever2011A, roever2011B}, where a filter was introduced based on {\it Student's t-distribution}, which can be proven to be robust against data non-stationarities. One can start from Eq.~(\ref{eq:gaussian}) and adopt a prior for the variance of the noise that follows the $\mrm{Inv}-\chi^2$ distribution. Then, following~\cite{roever2011A}, we can compute the marginal posterior by integrating out the noise variance, and arrive at the desired probability density function of Student's t-distribution that is more heavy-tailed than the Gaussian case. Student's t-distribution can be tuned with the  degrees-of-freedom parameter $\nu$, which can either be chosen a priori  or estimated directly from the data. A similar strategy was followed in other works, where the problem of data outliers was tackled by adopting a composite models for the total likelihood. For example, in~\cite{PhysRevD.60.021101} a Gaussian distribution was used for the noise and a uniform model for the data bursts, while in~\cite{PhysRevD.65.122002}, the total likelihood was a Gaussian mixture. 

In the following subsection, we will discuss the generalized hyperbolic model, which, at the cost of adding extra dimensionality to the problem, offers a more generic framework to handle a variety of data-irregularity situations.

\subsection{The Generalized Hyperbolic distribution\label{sec:generalized_hyperbolic_distr}}

We can now attempt to negate the possible shortcomings of the Whittle approximation by using the family of $\cal GH$ distributions~\cite{Bianchi2020, Borak2011,Prause1997ModellingFD, Eberlein2002TheGH, Prause1999TheGH}. One of the advantages of this practice, is that we can arrive at virtually any distribution of the exponential family, simply by tuning the parameters of the $\cal GH$ function. In fact, the Student's t-distribution mentioned above is a special case of the $\cal GH$ function, which, apart from the Student's t-distribution, leads to a large number of limiting distributions. The $\cal GH$ family of distributions for a variable $x$ can be expressed as:
\begin{equation}
\begin{array}{r@{}l}
    {\rm \cal GH} &{} (x | \lambda, \alpha, \beta, \delta, \mu) = \\
    &{} a(\lambda, \alpha, \beta, \delta, \mu) \left( \delta^2 + (x - \mu)^2 \right)^{(\lambda - \frac{1}{2})/2} \\
    &{} \times K_{\lambda - 1/2} \left( \alpha \sqrt{\delta^2 + (x - \mu)^2} \right) \exp\left[\beta (x - \mu)\right],
\end{array}
\label{eq:genhyp}
\end{equation}
where 
\begin{equation} 
a(\lambda, \alpha, \beta, \delta, \mu) = \frac{(\alpha^2 - \beta^2)^{\lambda/2}}{\sqrt{2\pi} \alpha^{\lambda - 1/2} \delta^\lambda K_\lambda \left(\delta \sqrt{\alpha^2-\beta^2} \right)},
\label{eq:a}
\end{equation}
and $K_\lambda$ is the modified Bessel function of the third kind. {\color{black} The domain of variation of the parameters $(\lambda, \alpha, \beta, \delta, \mu)$ is $\mu \in \mathbb{R}$ and
\begin{equation}
\begin{array}{ll}
\delta \geq 0,|\beta|<\alpha, & \text { if } \lambda>0, \\
\delta>0,|\beta|<\alpha, & \text { if } \lambda=0, \\
\delta>0,|\beta| \leq \alpha, & \text { if } \lambda<0 .
\end{array}
\label{eq:domain}
\end{equation}
Eq.~(\ref{eq:genhyp}) describes a 
skewed distribution when $\beta\neq 0$ and a symmetric distribution when $\beta = 0$.} The  parameter $\mu$ tunes the position of the distribution along the $x$-axis. {\color{black} The analytic expression for the variance is \cite{Prause1999TheGH}
\begin{equation}
    \sigma^2 = \frac{\delta K_{\lambda+1}(\delta\gamma)}{\gamma K_{\lambda}(\delta\gamma)} +\frac{\beta^2\delta^2}{\gamma^2}\Bigg[\frac{K_{\lambda+2}(\delta\gamma)}{K_{\lambda}(\delta\gamma)}-\frac{K_{\lambda+1}^2(\delta\gamma)}{K_{\lambda}^2(\delta\gamma)}\Bigg], 
\label{eq:variance}
\end{equation}
with $\gamma \equiv \sqrt{\alpha^2-\beta^2}$.  The general expression for the mean of the $\cal GH$ distribution is \cite{Prause1999TheGH}
\begin{equation}
    {\rm mean}(\cal GH) = \mu + \frac{\beta\delta}{\gamma} \frac{K_{\lambda+1}(\delta\gamma)}{ K_{\lambda}(\delta\gamma)}. 
\label{eq:mean}
\end{equation}
The ${\cal GH}$ distributions have semi-heavy tails. In particular, for $\mu=0$, the asymptotic behavior for x $\rightarrow \pm \infty$ is
\begin{equation}
{\cal GH}(x | \lambda, \alpha, \beta, \delta) \sim|x|^{\lambda-1} \exp ((\mp \alpha+\beta) x),
\end{equation}
up to a multiplicative constant \cite{BarndorffNielsen1981HyperbolicDA}. }

We can now define, as in~\cite{Prause1999TheGH}, the {\it {\color{black}multivariate, $d$-dimensional}} $\cal GH$ distribution for $\bm{x}\in {\rm I\!R}^d$, with $d$ being the given dimensionality. {\color{black}In the case of a GW detector  network, the dimensionality corresponds to the number of detectors. Then,
\begin{align}
{\rm \cal GH}_{d} (\bm{x} | \lambda, & \alpha, \bm{\beta}, \delta, \bm{\mu}) = \nonumber \\ 
& A \frac{K_{\lambda - d/2} \left( \alpha \sqrt{\delta^2 + r}\right)}{\left( \alpha^{-1}\sqrt{\delta^2 + r}\right)^{d/2 - \lambda}} \exp \left[ \bm{\beta}^{\rm T} (\bm{x}-\bm{\mu})\right],
\label{eq:genhypmd}
\end{align}
with $\bm{\mu}, \bm{\beta} \in {\rm I\!R}^d$, 
\begin{equation} 
A \equiv A(\lambda, \alpha, \bm{\beta}, \delta, \bm{\mu}) = \frac{\left( \sqrt{\alpha^2 - B}/\delta\right)^d}{(2\pi)^{d/2} K_\lambda \left( \delta \sqrt{\alpha^2 - B}\right)},
\label{eq:amd}
\end{equation}
where we set
\begin{equation}
    r = (\bm{x} - \bm{\mu})^{\rm T} \Delta^{-1} (\bm{x} - \bm{\mu}) , 
\label{eq:res}
\end{equation}}
\noindent and $B = \bm{\beta}^{\rm T} \Delta^{-1}\bm{\beta}$. In (\ref{eq:res}),  $\Delta$ is a positive definite matrix $\in {\rm I\!R}^{d\times d}$ with $\abs{\Delta} = 1$.

{\color{black}The $\cal GH$ distribution can be quite flexible in the description of data,
albeit at the cost of a larger number of parameters.
For a given problem with $d$ dimensions, the number of parameters is $2 d + 3$.}

\subsection{Limiting cases and subclasses of the Generalized Hyperbolic distribution}
\label{sec:limiting}

{\color{black}
Several well-known distributions are limiting cases or subclasses of the $\mathcal{GH}$ distribution. Below, we list some cases that we are going to use to set up test cases in Section \ref{sec:test-cases}.

\begin{itemize}
    \item The {\it Normal Gaussian} distribution ${\cal N}(\mu, \sigma^2)$ is a limiting case of the $\cal GH$ distribution, since
 \begin{eqnarray}
     {\mathcal N} &\leftarrow& \mathcal{GH}( \lambda=1, \beta=0, \alpha\rightarrow \infty, \delta \rightarrow \infty, \mu),
 \end{eqnarray}
provided that $\alpha \rightarrow \infty$ and $\delta \rightarrow \infty$ in such a way that $\delta / \alpha \rightarrow \sigma^2$.

 \item The {\it Normal Inverse Gaussian} distribution $\mathcal{NIG}(\alpha, \beta, \delta, \mu)$ has the explicit form \cite{barndorff1977exponentially}
    \begin{eqnarray}
        {\rm \cal NIG} {} (x | \alpha, \beta, \delta, \mu)  & = & \frac{\alpha\delta K_1 (\alpha\sqrt{\delta^2 + (x - \mu)^2}) }{\pi \sqrt{\delta^2 + (x - \mu)^2}} \nonumber\\
    && \times\exp{\big[\delta\gamma + \beta(x-\mu)\big]},
    \label{eq:nig}
    \end{eqnarray} 
which is a subclass of the $\mathcal{GH}$ distribution  for $\lambda=-1/2$
\begin{eqnarray}
    \mathcal{NIG}( \alpha, \beta, \delta, \mu) &=&  \mathcal{GH}( \lambda=-1/2, \alpha, \beta, \delta, \mu).
\end{eqnarray}
\item The {\it Hyperbolic} distribution $\cal H$ is a subclass of the $\mathcal{GH}$ distribution for $\lambda=1$
\begin{eqnarray}
    \mathcal{H}(\alpha, \beta, \delta, \mu) &=&  \mathcal{GH}(\lambda=1, \alpha, \beta, \delta, \mu),
    \label{eq:hyper}
\end{eqnarray}
whith $\delta \geq 0$ and $|\beta|<\alpha$. For $\beta=0$ it becomes the \textit{Symmetric Hyperbolic distribution}  ($\mathcal{SH}$). More generally, in the multivariate, $d-$dimensional case, the {\it multivariate Hyperbolic} (${\cal H}_d$) distribution is as subclass of the ${\rm \cal GH}_{d}$ distribution for $\lambda = (d + 1)/2$
\begin{eqnarray}
    \mathcal{H}_d(\alpha, \beta, \delta, \mu) &=&  \mathcal{GH}_{d}\left(\lambda=\frac{d + 1}{2}, \alpha, \beta, \delta, \mu \right),
\end{eqnarray}
\item The \textit{Student's t-distribution} with $\nu$ degrees of freedom is a limiting case of the $\mathcal{GH}$ distribution when $\lambda=-\nu/2, \alpha= 0, \beta= 0, \delta=\sqrt{\nu}$ and $\mu=0$.
\end{itemize}

 In Appendix \ref{sec:shapetriangle}, we demonstrate that one can recover the theoretical distributions of the above subclasses and limiting cases for test data, using Bayesian inference with appropriate likelihood functions.
 
We refer the reader to ~\cite{ernst, Bianchi2020, Borak2011,Prause1997ModellingFD, Eberlein2002TheGH, Prause1999TheGH} and references therein for a more extensive list of limiting distributions and subclasses of the $\cal GH$ distribution.
}

\subsection{The Hyperbolic likelihood function}
\label{seq:likelihood}

{\color{black}
Our main aim is to use a likelihood distribution that is sufficiently flexible, as to be able to reconstruct a large variety of distributions of data. As we demonstrate in the different examples in the next Sections, we find that a likelihood based on the multivariate Hyperbolic ($\cal{H}_{\rm d}$) distribution (i.e. the subclass of the $\cal{GH}_{\rm d}$ distribution when $\lambda = (d + 1)/2$) serves our purpose well. 

For a distribution of data $\bm{x}_i \in \mathbb{R}^d, 1 \leq i \leq n$, we derive the corresponding Hyperbolic likelihood function for inferring the parameters $(\alpha, \delta, \bm{\beta})$ of their distribution as\footnote{Our result reduces to the hyperbolic distribution $\Lambda_{\rm hyp}$ in \cite{Prause1997ModellingFD,Prause1999TheGH} for $\beta=0$.}} 
\begin{equation}
\begin{aligned} 
\Lambda_{\cal H}(\alpha, \delta, \bm{\beta}) = & n\Bigg[ \frac{d+1}{2} \ln\left(\frac{\gamma}{\delta}\right) + \frac{1-d}{2} \ln(2\pi)  \\
    & - \ln(2\alpha)  - \ln \left[K_{(d+1)/2} (\delta\gamma)\right]\Bigg] \\ 
    & - \alpha\sum^n_{i=1}\sqrt{\delta^2 + r_i}+ \bm{\beta}^{\rm T} \sum^n_{i=1}\bm{x}_i,
\label{eq:hyp_general}
\end{aligned}
\end{equation} 
{\color{black}where
\begin{equation}
r_i=\bm{x}_i^{\rm T} \widehat{\Delta}^{-1} \bm{x}_i.
\end{equation} 
We are going to use this form of the likelihood in test cases in Sec. \ref{sec:apps_and_usage}, where, for $d=1$, $r_i$ reduces to $r_i = x_i^2$.

For more realistic applications, we will adopt the above above Eq.~(\ref{eq:hyp_general}) in the frequency domain. In particular, in section~\ref{sec:gws} we adopt the {\it symmetric} Hyperbolic likelihood (i.e. $\bm{\beta}=0$). For multiple data channels - see Eq.~(\ref{eq:inprod}) - and assuming that $(\alpha, \delta)$ are the same for all frequencies, we then write

\begin{equation}
\begin{aligned} 
\Lambda_{\cal H}(\param, \alpha, \delta) = & n \Bigg[ \frac{d+1}{2} \ln\left(\frac{\alpha}{\delta}\right) + \frac{1-d}{2} \ln(2\pi)  \\
    & - \ln(2\alpha)  - \ln \left[K_{(d+1)/2} (\delta\alpha)\right]\Bigg] \\ 
    & - \alpha  \sum_i\sqrt{\delta^2 +  
{\rm Re}\, \left\{ r_i(\bm{\theta}) \right\}}, \\
\label{eq:hypf_whitened}
\end{aligned}
\end{equation} 
where
\begin{equation}
r_i(\bm{\theta}) = \tilde{\bm{\chi}}_i^\dag \mathbf{S}_n^{-1} \tilde{\bm{\chi}}_i .
\label{eq:ritheta}
\end{equation}
In the above relation, we have used \begin{equation}
\tilde{\bm{\chi}}_i = \sqrt{2 \mathrm{d}f}\left(\tilde{{\bm y}}_i-\tilde{{\bm h}}_i(\boldsymbol{\theta})\right),
\end{equation}
where $\mathrm{d}f$ is the given frequency resolution. The summation in Eq.~(\ref{eq:hypf_whitened}) is over elements that correspond to a chosen frequency range ($f_{\mrm{min}},\,f_{\mrm{max}}$).

Eq.~(\ref{eq:ritheta}) implies that we have chosen a given estimate (or model) for the PSD of the noise, through $\mathbf{S}_n$. One of the great advantages of this likelihood formulation, as demonstrated in the following sections, is that a mis-modeling of the noise PSD could be compensated by the joint estimation of the $(\alpha, \delta)$ parameters. These two parameters will also indicate possible departures from a Gaussian distribution, as described in section~\ref{sec:limiting}.
In the rather exceptional case where one does not want to assume any PSD model, but instead prefers to infer the noise properties by using the $(\alpha, \delta)$ parameters of the Hyperbolic likelihood, then one can use 
\begin{equation}
r_i(\bm{\theta}) =   \tilde{\bm{\chi}}_i^\dag {\widehat{\Delta}}^{-1} \tilde{\bm{\chi}}_i ,
\label{eq:ri_nonwhite}
\end{equation}
where now, since we operate in the frequency domain, and in order to arrive at a dimensionless $r_i$, the $\widehat{\Delta}$ has units of $1/\mathrm{Hz}$ (alternatively, one could make ${\tilde {\bm \chi}}_i$ dimensionless, by scaling it with an appropriate reference value with units of 1/$\sqrt{\rm Hz}$).} 

After defining the likelihood function, and since we operate within a Bayesian framework, we can write the posterior of the parameters as 
\begin{equation}
    p(\param | {\bm y}) \propto p({\bm y}|\param)p(\param),
    \label{eq:bayes}
\end{equation}
with $p(\param)$ being the prior of the parameters. The marginal likelihood $p({\bm y})$ usually acts as a normalization constant and for that reason it is omitted from Eq.~(\ref{eq:bayes}). \\


{\color{black}

\subsection{Test cases\label{sec:apps_and_usage}}

Before applying the hyperbolic likelihood $\Lambda_{\cal H}$ to examples in GW astronomy, we first demonstrate its ability to reconstruct of non-Gaussian test distributions (several analytic distributions, as well as real data from the LISA Pathfinder \cite{Armano:2016bkm, Armano:2018kix} mission).

\subsubsection{Reconstruction of one-dimensional test distributions}
\label{sec:test-cases}

In the following, we will reconstruct different one-dimensional test distributions (corresponding to the liming cases and subclasses of the $\cal GH$ distributions discussed in Section \ref{sec:limiting}), assuming that the $\cal H$ distribution can approximately describe the difference cases (even those with $\lambda \neq 1$). For each case, we infer the $(\alpha, \beta, \delta, \mu)$ parameters of $\cal H$ using a Metropolis-Hastings MCMC algorithm and $\Lambda_{\cal H}$ (with $d=1$) as the likelihood function.} The purpose of this first simple investigation is to demonstrate the flexibility of the $\Lambda_{\cal H}$ likelihood. 

For the sake of simplicity, we generate~\footnote{The random values are generated with the \texttt{scipy.stats} package~\cite{SciPy}.} all data {\color{black}($5\times10^5$ samples for each test distribution)} with zero-mean ($\mu=0$). {\color{black}The detailed description of each test distribution is as follows:  

\begin{enumerate}
    \item Normal Gaussian distribution $\mathcal{N}(\mu=0,\sigma^2=1)$. 

    \item Normal Inverse Gaussian distribution ${\cal NIG}( \alpha=1, \beta=0, \delta=0.8, \mu=0)$. 

    \item Hyperbolic distribution, ${\cal H}(\alpha=1.5, \beta=0.75, \delta=2,\mu=0)$.
  
    \item Student's t-distribution with $\nu=4$.

\end{enumerate}

}
    
\begin{figure}
         \includegraphics[width=0.48\textwidth]{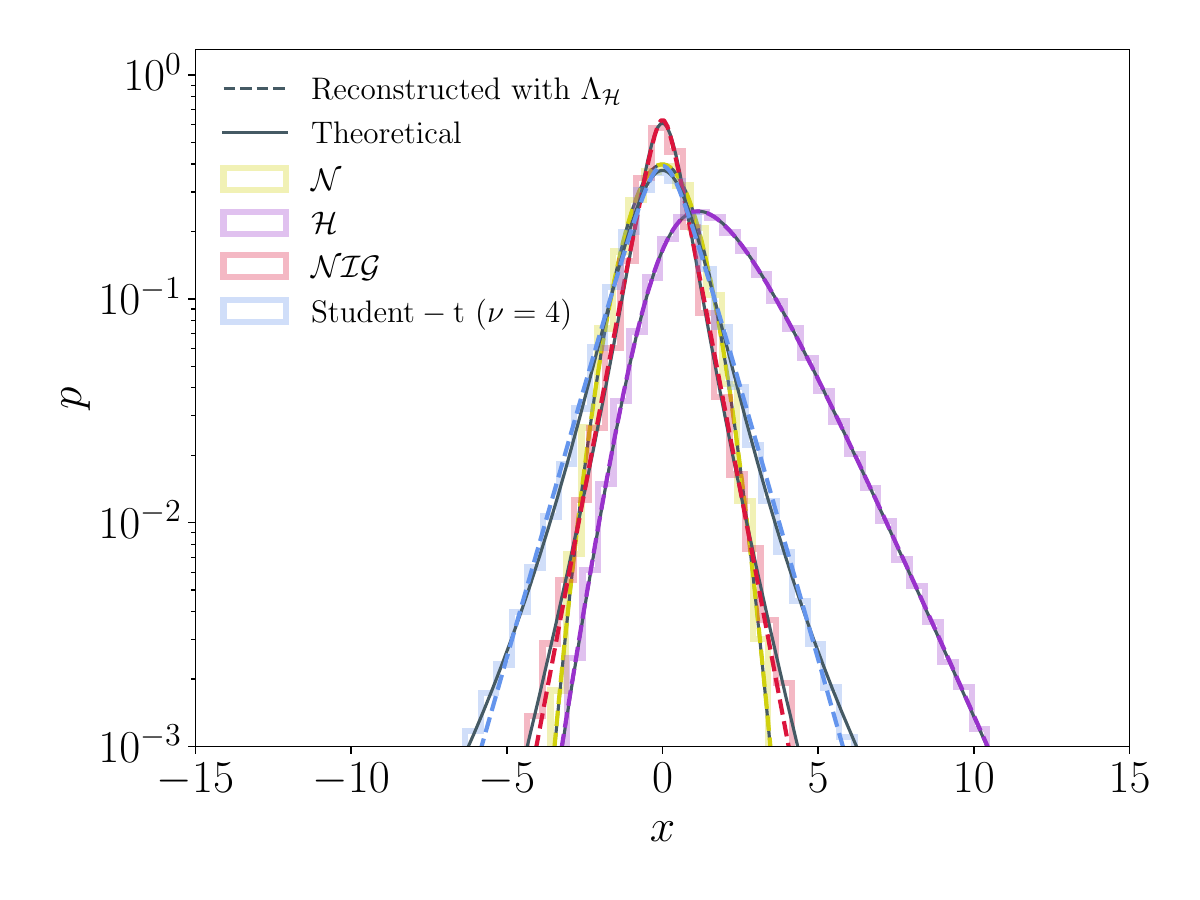}
        \caption{Randomly generated data with Gaussian $\mathcal{N}$ (yellow),  hyperbolic $\mathcal{H}$ (purple), $\cal NIG$ (red), and Student's t (light blue) distributions.  {\color{black}The reconstructed distributions, using Bayesian inference with $\Lambda_{\cal H}$ as the likelihood function (solid lines), agree very well with the theoretical distributions (dashed lines).}}
        \label{fig:toy_distributions}
\end{figure}

{\color{black}The generated test distributions described above are shown in Figure~\ref{fig:toy_distributions}. The solid lines represent the theoretical distribution in each case. The dashed lines represent our reconstruction using the $\Lambda_{\cal H}$ likelihood for all cases. We see that the two cases for which $\lambda=1$ ($\cal N$ and $\cal H$) are reconstructed with high accuracy (the Jensen-Shannon divergence\footnote{\color{black}The Jensen-Shannon divergence~\cite{MENENDEZ1997307, nielsen2020generalization} is a special-case of the Kullback–Leibler divergence, which can be used as a metric for the similarity of two distributions. It is symmetric, and takes values close to zero if two distributions are similar.} is smaller than $10^{-5}$). But also, the two cases with $\lambda \neq 1$ ($\cal NIG$ and Student's t) are still reconstructed fairly accurately (the Jensen-Shannon divergence is $2.5\times 10^{-4}$ and $5.1\times 10^{-4}$ respectively) when using the same $\Lambda_{\cal H}$ likelihood. This demonstrates the flexibility of the $\Lambda_{\cal H}$ likelihood to reconstruct different distributions of data and to compensate for different underlying values of $\lambda$ through its other parameters.

Finally, one can use the $\{\alpha,\,\beta,\,\delta\}$ parameters to probe the statistical properties of the residuals. For example, in many cases, it is necessary to test for any departures from Gaussianity for a given data set. A very useful graphical tool for such applications is the so-called {\it shape triangle}, where, based on the recovered $\{\alpha,\,\beta,\,\delta\}$ coefficients, we can qualitatively characterize the yielding distribution of the residuals~\cite{Prause1999TheGH,KUCHLER19991}. We present more details about this methodology in the Appendix~\ref{sec:shapetriangle}.
}

\subsubsection{Application to real data from the LISA Pathfinder mission}

LISA Pathfinder (LPF) was an ESA mission launched in late 2015 and remained in operation until 2017~\cite{Armano:2016bkm, Armano:2018kix}. The primary goal of LPF was to test technologies for future GW observatories in space, such as the LISA mission. LPF was essentially a laboratory in space, which contained two cubic test masses maintained in free-fall conditions. The differential acceleration $\Delta g (t)$ between the two test masses was being monitored by means of laser interferometry. During the mission, the differential acceleration noise budget was studied and modeled, with the aim of building solid grounds for the development of the LISA mission.
One of the most important measurements of the LPF mission was the pure acceleration noise measurements, where no experiments (excitations of the three-body system) were performed. During those measurements, a great variety of spurious  transient signals (glitches) were recorded. While the physical origin of those glitches is not entirely known~\cite{LISAPathfinder:2022awx}, their statistical properties were studied in detail~\cite{Baghi:2021tfd}. {\normalfont For more studies on the LPF acceleration noise data, we refer the reader to~\cite{Thorpe_2019, Armano:2018ucz, 2019PhRvD..99h2001A, 2018PhRvD..98j2005A, 2017PhRvL.118q1101A}.} 

\begin{figure}
    \centering
    \includegraphics[width=1.03\linewidth]{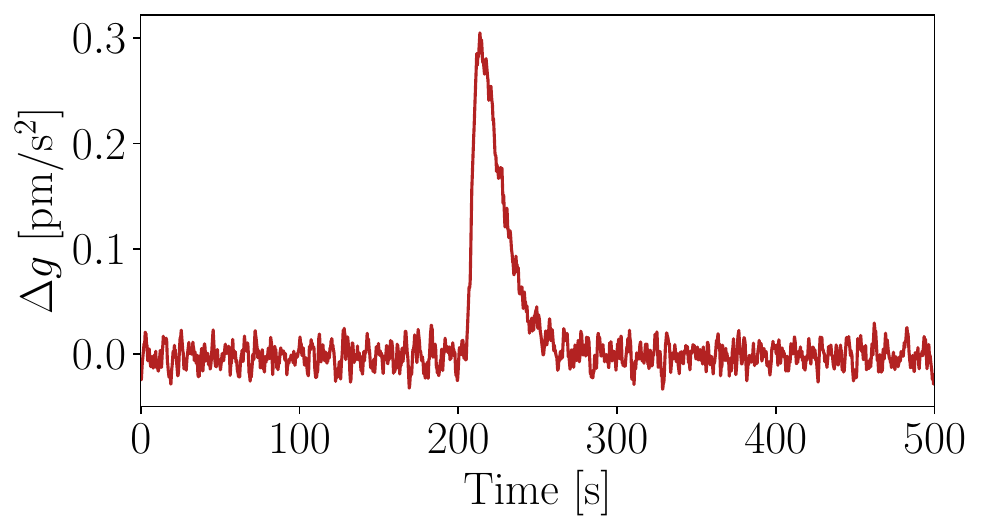}
    \includegraphics[width=1\linewidth]{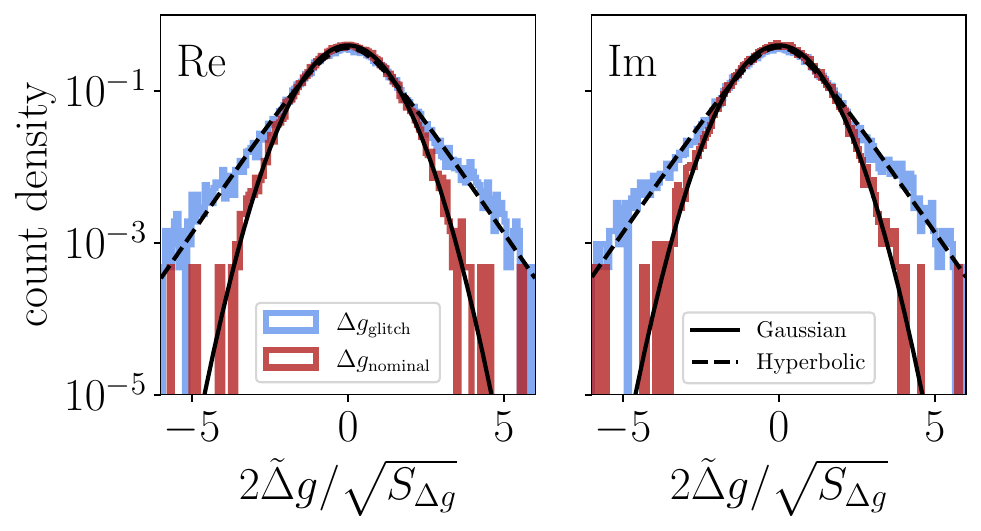}
    \caption{\textit{Top panel:} A loud glitch in the $\Delta g(t)$ time series data from the LPF mission (see text for details). The data presented here are de-trended {\normalfont by subtracting their mean and low-passed for visualization purposes}. \textit{Bottom panel:} 
    {\color{black} Histograms of the distributions $\tilde{\Delta} g$ in the frequency domain, normalized  with the PSD $S_{\Delta g}$ of} a nearby-in-time, outlier-free segment. \textit{Blue:} segment that contains the loud glitch. \textit{Red:} reference data set. The dashed and dotted lines represent Gaussian and Hyperbolic distributions, respectively. }
    \label{fig:lpf}
\end{figure}

For our example here, we take the $\Delta g(t)$ segment that was recorded between 2017-02-13 14:30:00.000 and 2017-03-02 21:50:19.000 UTC. This particular segment contained a series of glitches, with the loudest of them (shown in the top panel of Fig.~\ref{fig:lpf}) distorting the numerical estimations of the PSD of the data. In fact, such strong spurious signals were straightforwardly detected and subtracted from the data. However, this situation of data outliers is ideal for testing the heavy-tailed likelihood framework we introduced in Sec. \ref{seq:likelihood}. 

We first crop\footnote{{\normalfont We remove the first and last few data points to avoid distortions due to the application of filters.}} the $\Delta g(t)$ time series and determine the time of occurrence of the loudest glitch signals. We also mark the $\Delta g(t)$ segments when no glitches are present, with the aim of using them as benchmarks for the PSD of $\Delta g(t)$. Then, we proceed and apply the Hyperbolic likelihood {\color{black}function  Eq.~(\ref{eq:hyp_general}) for the distribution of the $\tilde{\Delta} g(f)$ data in the frequency domain and infer the free parameters $\alpha$, $\beta$, and $\delta$}. In order to avoid getting trapped in local maxima on the posterior surface, we have used MCMC methods~\cite{eryn}. 

We plot the results in the bottom panel of Figure~\ref{fig:lpf}, where the histograms of the distribution of the noise data points are plotted. The distribution of the $\Delta g$ noise for the particular segment that contains the loud glitch is represented in red color. For the sake of comparison, we have also chosen a neighboring-in-time segment, where no data outliers are present (blue data). The hyperbolic density which is computed at the estimated $\alpha$, $\beta$, and $\delta$ is shown with the dotted line, while the theoretical Gaussian distribution is represented with the dashed line.  {\color{black} The heavy-tailed hyperbolic distribution, Eq. (\ref{eq:hyp_general}), indeed manages to correctly capture the tails of the distribution in the case where a loud glitch exists in the data, whereas the Gaussian distribution is suitable only for the segment where no large glitch was detected. We note that, due to the nearly symmetric distribution of the data, the parameter $\beta$ was inferred to be very close to zero and a symmetric hyperbolic likelihood would still be a good choice.}

These first results are very encouraging, thus we can now proceed and apply this framework to more complicated investigations. Such cases, as already discussed, are situations of having to search for signals in noisy data, where it is crucial to be able to correctly model the statistical properties of the residuals and the underlying noise. In the following Sections, we will apply the {\color{black}hyperbolic likelihood function} in more realistic applications drawn from challenges in Gravitational Wave data analysis.

{\color{black}In this first study, we are going to focus on symmetric hyperbolic distributions (i.e. with $\beta = 0$) for two main reasons. Firstly, we have found that in the simple applications presented in this work, the symmetric Hyperbolic distribution is more than sufficient, and secondly, because it is  lighter in terms of computational requirements. Indeed, for a measurement of $n$ data points, the number of Bessel function calculations is reduced to one, instead of $n+1$ for the case of asymmetry, where $\beta \neq 0$.}


\section{Applications to the analysis of Gravitational Wave data\label{sec:gws}}

The detection and characterization of any given signal requires a good level of knowledge of the instrumental noise. Concerning the GW measurements from ground-based detectors, there has been extensive work on the statistical properties of the instrumental noise (e.g.~\cite{Martynov, Abbott_2020_noise,Virgo_noise} and references therein). This, together with monitoring the status of the instruments through a large number of auxiliary channels, contributes to the accurate representation of the instrument capabilities, virtually at all times.

However, future detectors, such as LISA, are going to be signal-dominated, and the noise of the observatory will not be completely known. In particular, LISA is going to measure the complete population of Ultra Compact Binaries emitting from within the Milky Way~\cite{Amaro2017}. These objects are mostly Double White Dwarfs, and their total number is estimated to be of order ${\mathcal O}(10^6)$, emitting nearly monochromatic GW radiation. Depending on the given population model, current estimates place the total  number of sources resolvable  by LISA at ${\mathcal O}(10^4)$, while the rest will generate a non-stationary confusion signal between about $0.01$ and $0.2$~$\mrm{mHz}$~\cite{karnesis21,georgousi22,korol22}. At the same time, the number of calibration instruments available on board a space-borne observatory (thermometers, magnetometers, etc.) will be fairly restricted,
due to space and power supply limitations. The above, in combination with  the limited data transmission rate, might contribute to having less information about the instrumental noise, when compared to ground-based observatories. We will utilize examples drawn from LISA data scenarios, focusing mostly on analyzing signals from UCBs. In the following Section, we will compare the performance of the different likelihood formulations described in Section~\ref{sec:background}, under different noise assumptions.

\subsection{Methodology \label{sec:method}}

In order to test the performance of the Hyperbolic likelihood, we designed a series of experiments with synthetic data. We chose to work only with the signals from UCBs, mostly due to their monochromatic nature, and also for practical reasons, such as the very small computational time. However, it should be mentioned here, that our formulation applies straightforwardly to the chirping signals of supermassive black hole binaries. For those signals, and the simplest noise case of a varying PSD level across the complete frequency band (same spectral shape, different overall amplitude), we obtain very similar results as with the UCBs investigations presented in this Section. The situation becomes more complicated, however, when the noise PSD level assumed in the likelihood function has a different spectral shape than the true noise PSD. We leave this kind of investigation to future work.

For our first experiment, we simulate data given a particular instrument sensitivity and then perform the analysis assuming {\color{black}fixed PSD} of the noise. We use three likelihood formulations. The first type refers to the standard Gaussian likelihood of Eq.~(\ref{eq:gaussian_llh}), but {\color{black} with an assumed noise PSD which is chosen to be different than the one that was used to generate the synthetic data}, simulating situations of instrumental noise mismodeling. The second type refers to the Whittle likelihood of Eq.~(\ref{eq:whittle_llh}), where the {\color{black}a model for the noise PSD is being fitted simultaneously with the GW waveform parameters}. Finally, the third refers to the hyperbolic likelihood of Eq.~(\ref{eq:hypf_whitened}), with {\color{black}fixed PSD for the noise as in the case of the Gaussian likelihood above, but with the added flexibility of extra parameters}. 

It is worth noting here, that for our first experiments, we have assumed ideal synthetic data. This means that we simulate perfectly Gaussian and uninterrupted data, directly from the LISA sensitivity curves~\cite{snrtn}. At the same time, we assume a rigid LISA constellation, while the test-masses acceleration noise and spacecraft interferometric noises are assumed to be equal for all three spacecraft~\cite{Prince2002}. For this special case, we can confidently utilize the noise-orthogonal Time Delay Interferometry (TDI)  $A,\,E$, and $T$ channels~\cite{Tinto:2004wu, snrtn}, which are retrieved from the detector outputs $X, \,Y, \, Z$ as
\begin{equation}
\begin{aligned}
    A = \frac{1}{\sqrt{2}}(Z - & X), \quad E = \frac{1}{\sqrt{6}}(X - 2Y + Z), \\
    T &= \frac{1}{\sqrt{3}}(X + Y + Z).
    \label{eq:aet}
\end{aligned}
\end{equation}
This also means that the $A$ and $E$ channels used in the analysis will have the same spectral shape, which under given assumptions about the future instrument, is given by~\cite{snrtn,scird}
{\color{black}
\begin{eqnarray}
S_n^{\mrm{A}\mrm{E}}(f) = & 8\, \mrm{sin}^2(f^\ast) \big[ 2  S^\mrm{acc}_{\delta\nu/\nu}(A;f)\left( 3 + 2\mrm{cos}(f^\ast) + \mrm{cos}(2f^\ast) \right) \nonumber\\
& + S^\mrm{OMS}_{\delta\nu/\nu}(P;f) \left( 2+\mrm{cos}(f^\ast)\right)  \big],
\label{eq:sim_psd_model}
\end{eqnarray}
where $f^\ast = 2 \pi Lf/c $, $L$ is the LISA arm-length, while the noise components $S^\mrm{acc}_{\delta\nu/\nu}(A;f)$ and $S^\mrm{OMS}_{\delta\nu/\nu}(P;f)$  can be found in~\cite{snrtn},  expressed in relative frequency units (thus the index $\delta\nu/\nu$). These functions depend on parameters $A$ and $P$, respectively, for which a conservative estimate consistent with current requirement levels for the instrument}  is~\cite{snrtn,scird,Flauger:2020qyi}
\begin{equation}
\begin{aligned} 
\sqrt{A} = & 3~\mrm{fm / sec^2 / \sqrt{Hz} } , \; \text{and} \\
\sqrt{P} = & 15~\mrm{pm/\sqrt{Hz}}.
\label{eq:sim_psd_params}
\end{aligned}
\end{equation}
The $T$ channel is the so-called {\it null} channel, which greatly suppresses the GW signal, and for that reason, it will not be considered in our analysis. {\color{black} The above assumptions reduce the likelihood computations of Eqs.~(\ref{eq:gaussian_llh}),~(\ref{eq:whittle_llh}), and~(\ref{eq:hypf_whitened}) to just the sum of the likelihoods for the two $A$ and $E$ channels, which under the assumptions above, and for our application in the following section, can be considered  uncorrelated.}

Finally, in order to perform our analysis, we sample the posterior of the parameters with Markov Chain Monte Carlo algorithms, enhanced with \textit{parallel tempering}~\cite{eryn}. For the applications in the Sections below, we have set up our sampler with an adjusting temperature ladder of $30$ temperatures, each running with $50$ independent walkers. We nominally allow for $5\times10^{4}$ samples for the burn-in phase, and $10^{5}$ samples per walker for the parameter estimation phase.

\begin{figure*}
     \centering
     \begin{subfigure}[b]{0.42\textwidth}
         \centering
         \includegraphics[width=\textwidth]{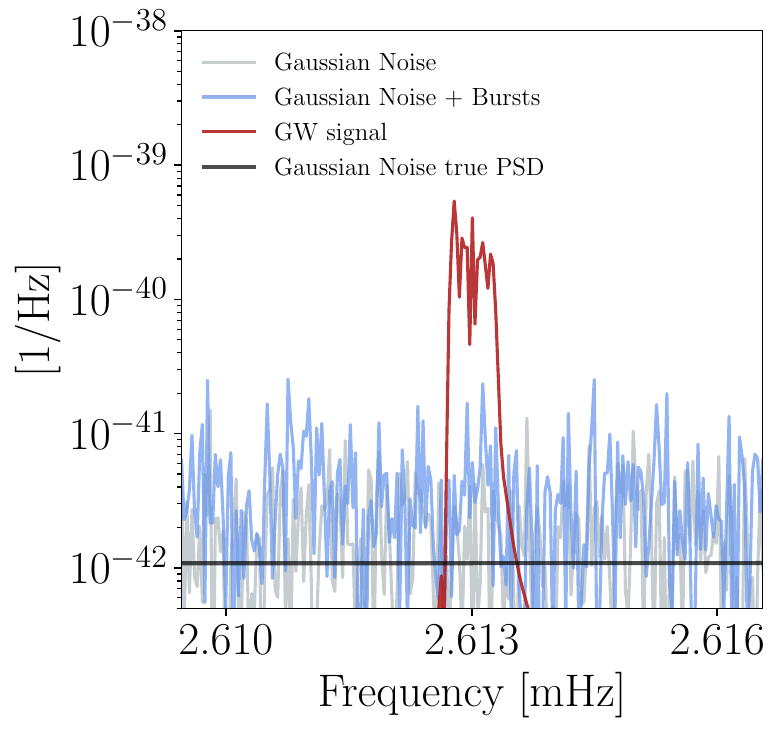}
         \caption{}
         \label{fig:ubcs_burst_data_freq}
     \end{subfigure}
     \quad\quad
     \begin{subfigure}[b]{0.4\textwidth}
         \centering
         \includegraphics[width=\textwidth]{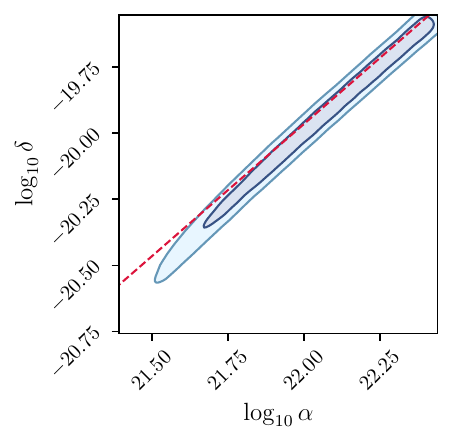}
         \caption{}
         \label{fig:ubcs_alpha_delta}
     \end{subfigure}
        \caption{\textit{(a)} {\color{black} The different sets of data used in the investigations of section~\ref{sec:gws}. In the given frequency range, the PSD of the instrumental noise is shown with the solid black line (see Eq.~(\ref{eq:sim_psd_model}), (\ref{eq:sim_psd_params})). This model, {\color{black}with added Gaussian noise} was used in order to generate the data shown in gray color. For the investigation in section~\ref{sec:ucbs_bursts}, we additionally injected a series of Gaussian bursts, which resulted in the noise data shown in blue color. The injected UCB signal is shown in dark red (see Table~\ref{tab:ucb_params}).} \textit{(b)} The $2\mrm{D}$ posterior slice between the parameters $\alpha$ and $\delta$  of the Hyperbolic likelihood defined in Eqs.~(\ref{eq:hypf_whitened}, \ref{eq:R_unwhiten}), for the first investigation of Section~\ref{sec:ucbs_level}. The contours represent the sampled posterior surface {\color{black}at 67\% and 90\% C.I.}, while the dashed red line corresponds to the logarithm of the true injected PSD levels of the instrumental noise (see text for details).}
        \label{fig:ucbs_noise_fit}
\end{figure*}

\subsection{Ultra Compact Galactic Binaries \label{sec:gbs}}

As already discussed, the UCBs emitting from the vicinity of our Galaxy are mostly Double White Dwarfs (DWDs)~\cite{korol22,Stephan_2019, Korol_Toonen}. Those will be the most numerous types of sources in the LISA band, with millions of them emitting in the $\mrm{mHz}$ range. For our study here, we will perform parameter estimation on a single binary embedded in Gaussian noise, \color{blue} with a PSD given by Eq. (\ref{eq:sim_psd_model})}. For the actual waveforms, we utilize the fast frequency-domain  implementation that was first presented in~\cite{cornish07}. The two polarizations are then written as
\begin{equation}
\begin{aligned} 	
    h_+(t)& = \frac{2\mathcal{M}}{D_L}\left(\pi f_0(t)\right)^{2/3}\left( 1 + \cos^2 \iota \right) \cos{\psi(t)}, \\
	h_\times(t)& = -\frac{4\mathcal{M}}{D_L}\left(\pi f_0(t)\right)^{2/3}\cos{\iota} \sin{\psi(t)}, \label{eq:ucbhphc}
\end{aligned}
\end{equation}
with $\mathcal{M}$ being the chirp mass, $f_0$ the instantaneous gravitational wave frequency, $D_L$ the luminosity distance, $\iota$  the inclination of the binary orbit and $\psi$ the gravitational wave phase over time. For more details about the waveform model, we refer the reader to~\cite{robson18, cornish07, katz_exo22}. The final inferred parameters are  $\param = \{ \log_{10}\mathcal{A},\, f_0\, \log_{10}\dot{f}_\mrm{gw},\, \phi_0,\, \cos\iota, \psi, \lambda_{\rm E}, \sin\beta_{\rm E} \}$, where $\mathcal{A}$ is the overall amplitude, which, in relation to Eqs.~(\ref{eq:ucbhphc}), can be expressed as
\begin{equation}
	\mathcal{A} = D_L^{-1}\left(2 \mathcal{M}^{5/3}\pi^{2/3} f_0^{2/3} \right) .
	\label{eq:amplitude}
\end{equation}
Above, $\lambda_{\rm E}$ and $\beta_{\rm E}$ are the given ecliptic latitude and longitude, respectively. The GPU-accelerated software we have used can be found in~\cite{katzucbszenodo}, which is essentially an adaptation of the waveform implemented for the LDC software~\cite{ldcsoft} and in earlier works. The injected parameters for the single binary are presented in Table~\ref{tab:ucb_params}.
\begin{ruledtabular}
\begin{table}
\centering
    \begin{tabular}{l@{\hskip 0.05in}c}
       \textbf{Parameters} &\textbf{Value}  \\
      \hline
    Amplitude, $\log_{10}(\mathcal{A}~[\mrm{Strain}])$ & -21.37729133  \\ 
    Initial frequency, $f_0~[\mrm{mHz}]$ & 2.61301   \\
    Frequency derivative, $\log_{10}\dot{f}_0$ & -16.53675992  \\
    Inclination, $\cos(\iota~[\mrm{rad}])$  & 0.05407993  \\ 
    Ecliptic Latitude, $\sin(\beta_{\rm{E}}~[\mrm{rad}])$ & 0.1012303 \\
    Ecliptic Longitude, $\lambda_{\rm{E}}~[\mrm{rad}]$ & 4.052962883  \\
    Polarization, $\psi~[\mrm{rad}]$ & 0.80372815 \\
    Phase, $\phi_0~[\mrm{rad}]$ & 3.12184095 \\
    \end{tabular}
    \caption{The parameter values of the injected UCB signal in the data. This injection was used for the investigations in both Sections~\ref{sec:ucbs_level} and~\ref{sec:ucbs_bursts}.  \label{tab:ucb_params}}
\end{table}
\end{ruledtabular}

\subsubsection{Unknown noise spectrum \label{sec:ucbs_unknown_noise}}

{\color{black} We first investigate the case, where the true PSD of the noise is completely unknown. A partial solution for this problem would be to assume a parameterized model of the noise PSD and infer its parameters simultaneously with the signal, using the Whittle likelihood of Eq.~(\ref{eq:whittle_llh}). Here, instead, we will use the Hyperbolic likelihood $\Lambda_{\cal H}$ of Eq.~(\ref{eq:hypf_whitened}), with ${\bm r}_i(\bm{\theta})$ given by Eq. (\ref{eq:ri_nonwhite}), without making an explicit assumption about the noise PSD.

By adopting the noise-orthogonal $A$ and $E$ channels, Eq.~(\ref{eq:ri_nonwhite}) reduces to
\begin{equation}
    r_i(\bm{\theta}) = 2 \mathrm{d}f \sum_c^{\{ A, E\}} \abs{\tilde{y}_{c} - \tilde{h}_{c}(\param)}^2.
\label{eq:R_unwhiten}
\end{equation}
Essentially, this corresponds to using a unitary diagonal $\hat \Delta$ matrix.} 

Given the simulated data, we should recover a set of $\alpha$ and $\delta$ parameters that would reflect the 
{\color{black}characteristics of the noise. Since, in  our example, we generated Gaussian noise, the posterior distribution of $\alpha$ and $\delta$ should reflect the characteristics of the Gaussian normal distribution that we have adopted for the simulations. }

For this investigation, we can treat the instrumental noise as white around the injection frequency, simply because this type of signal is almost monochromatic and, thus, the variation of the noise PSD {\color{black}in a narrow frequency band} around the signal is negligible (see Figure~\ref{fig:ubcs_burst_data_freq}). This further simplifies our analysis because it allows us to use the common $\alpha$ and $\delta$ parameters for all frequencies considered in the analysis. In {\color{black}other cases, such as when one simulates multiple injections at different frequencies,
different $\alpha, \delta$ parameters will need to describe different parts of the spectrum}
as demonstrated {\color{black} in an example using Student's t-distribution} in~\cite{roever2011B}. Another simplification is that we use the same $\alpha$ and $\delta$ coefficients for both the  $A$ and $E$ TDI variables, assuming that the two data channels have the same noise levels\footnote{ {\color{black} For future applications, it would be interesting to generalize the $\cal{GH}$ distribution to $\alpha, \delta \in \mathbb{R}^d$.}}.

Considering the above, we should expect to recover parameters that while $\alpha \rightarrow \infty$ and $\delta \rightarrow \infty$, their ratio remains constant and converges to the variance of the noise around the source initial emission frequency $f_0$. Concerning the signal, 
we draw a relatively ``loud'' source from the list of Verification Binaries~\cite{ldc, ldcsoft}, which we further tune to get a slightly higher SNR. The injected waveform parameters are listed in Table~\ref{tab:ucb_params}, while our  synthetic data set has a duration of $T_\mathrm{obs} = 1~\mathrm{year}$, {\normalfont which gives a frequency resolution of $\sim10^{-8}~\mrm{Hz}$}.

We then perform parameter estimation {\color{black}simultaneously for the waveform parameters listed in Table~\ref{tab:ucb_params} and the parameters $\alpha, \delta$ of the $\Lambda_{\cal H}$ likelihood}. {\color{black}Figure~\ref{fig:ubcs_alpha_delta} displays the 2D posterior distribution of the parameters $\alpha, \delta$. For a symmetric distribution, Eq. (\ref{eq:variance}) reduces to $\sigma^2 \rightarrow \delta/\alpha$ in the limit of a Gaussian distribution\footnote{{\color{black}Notice that, as evident from Figure~\ref{fig:ubcs_burst_data_freq}, $S_n^{\rm AE}(f_0)\sim 10^{-42} {\rm Hz}^{-1}$, which explains the particular numerical values of $\alpha$, $\delta$ in the chosen system of units, where $\hat \Delta$ is assumed to have dimensions of $1/{\rm Hz}$ without any further scaling of the ${\tilde {\bm\chi}}_i$ data. With appropriate rescaling, one can bring both $\alpha$, $\delta$ in the regime of $\rightarrow \infty$.}}, i.e. when $\alpha\rightarrow \infty$ and $\delta\rightarrow \infty$.
The dashed red line in Figure~\ref{fig:ubcs_alpha_delta}  represents $\log_{10}\delta - \log_{10}\alpha = \log_{10}\sigma^2$, where  $\sigma^2$ here corresponds to the ``correct" value of the power spectral density $S_n^{\mrm{A}\mrm{E}}(f_0)$ in the same dimensionless units we chose for the data. From the alignment of the 2D posterior distribution with this line, it follows that the inferred parameters $\alpha, \delta$ adjust the Hyperbolic distribution to the true Gaussian distribution of the data residuals.} 

Finally, the waveform parameters are also recovered with identical posteriors as the baseline Gaussian likelihood case, where the noise in Eq.~(\ref{eq:gaussian_llh}) is set to the true PSD value. This first result demonstrates the capabilities and robustness of the Hyperbolic likelihood in recovering both the correct parameters of the noise and the statistical properties of the residuals. 
\begin{figure*}
 	\includegraphics[width=.9\linewidth]{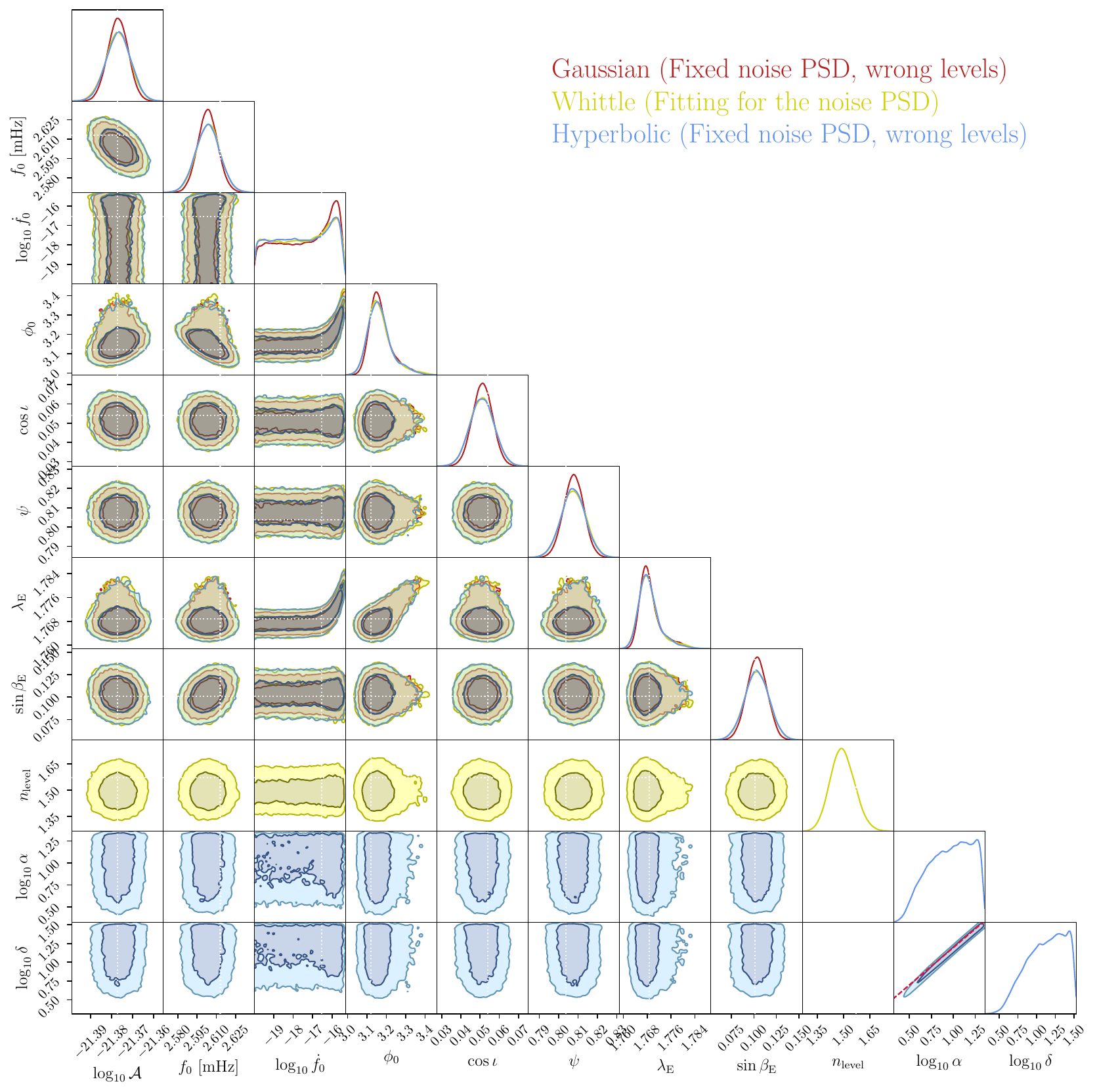}
 	\caption{ Corner plot showing the 2D posterior slices for all the parameters {\color{black}that characterize the injected waveform and the properties of the noise}, for the case of noise mismodeling of Section~\ref{sec:ucbs_level}. {\normalfont The optimal signal-to-noise ratio is $122.7$}. In particular, with dark red we plot the posteriors for the case of the Gaussian likelihood $\Lambda_{\cal N}$ with {\color{black} PSD at a wrong level}, whereas with yellow we plot the posteriors of the Whittle likelihood ($\Lambda_{\cal W}$), which allows for fitting of the noise level. Finally, with light blue, we plot the posteriors retrieved with the Hyperbolic likelihood, with the wrong PSD levels in the whitening. The true injected values for all parameters (with the exception of $\alpha$ and $\delta$ for the Hyperbolic likelihood case) are shown with white dashed lines. The red dashed curve represents the $\log_{10}\delta - \log_{10}\alpha = \log_{10}n_\mrm{level}^{(\mrm{correct})}$ line. }
\label{fig:ucbs_triangle_level}
\end{figure*}

\subsubsection{Noise spectral model mismodeling \label{sec:ucbs_level}}

Our next step is to test this framework in a different setting, which focuses on mismodeling the PSD of the noise. {\color{black}This situation is a bit closer to reality, where we start making some assumptions about the underlying PSD of the noise.} 
{\normalfont This will be the case for LISA around the~$\mathrm{mHz}$ level, due to the unknown contribution of the stochastic signal generated by astrophysical populations~\cite{Babak:2023lro, Pozzoli:2023kxy,karnesis21,korol22,LISACosmologyWorkingGroup:2022jok}.}

In practice, we simulate a mismodeling situation by plugging a wrong PSD model of the noise to the likelihood function of Eq.~(\ref{eq:gaussian_llh}), (\ref{eq:whittle_llh}) and~(\ref{eq:hypf_whitened}). Since our signal is almost monochromatic, this translates to an overall PSD level difference between the true and adopted models for the noise. We use the same source signal as before (see Table~\ref{tab:ucb_params}), and simulate data for an observation duration of $T_\mathrm{obs} = 1~\mathrm{year}$. {\color{black} Using a PSD based on Eq. (\ref{eq:sim_psd_model}) and assuming that the values of Eq. (\ref{eq:sim_psd_params}) correspond to the ``correct" noise levels, we obtain a signal-to-noise ratio (SNR) of $\rho =\sqrt{\langle h|h\rangle} = 122.7$. 

Next, we assume that we do not know the ``correct" noise levels and are forced to guess, adopting a different, ``wrong" PSD, using Eq. (\ref{eq:sim_psd_model}), but with the ``wrong" values}, 
\begin{equation}
\begin{aligned} 	
    \sqrt{A} &= 2.5~\mrm{fm / sec^2 / \sqrt{Hz}}, \; \text{and} \\
    \sqrt{P} &= 11~\mrm{pm/\sqrt{Hz}},
    \label{eq:sim_psd_params_wrong}
\end{aligned}
\end{equation}
which means that by adopting the above model, we greatly underestimate the level of the noise by a factor of 
$\sim 1.57$ at the particular emission frequency of the binary. We then sample the parameter space with parallel tempering MCMC~\cite{eryn} {\color{black} and investigate three different cases}.
\begin{enumerate}
    \item The {\it first} parameter estimation analysis was performed by adopting the standard Gaussian likelihood ($\Lambda_{\cal N}$) of Eq.~(\ref{eq:gaussian_llh}), and assuming the ``wrong" PSD levels of Eq. (\ref{eq:sim_psd_params_wrong}) for the noise. 

    \item The {\it second} investigation was analyzed using the Whittle likelihood ($\Lambda_{\cal W}$) of Eq.~(\ref{eq:whittle_llh}), {\color{black}with a model of the PSD that represents a flat PSD spectrum around the frequencies of interest ($f_{\rm min}= f_0 - 1~[\mathrm{mHz}], f_{\rm max}= f_0 + 1~[\mathrm{mHz}]$) multiplied by a single parameter $n_{\rm level}$. For $n_{\rm level}=1$, we take the PSD to be the value given by Eq.(\ref{eq:sim_psd_model}) at $f=f_0$, but in which the ``wrong'' parameters of Eq. (\ref{eq:sim_psd_params_wrong}) were used. By fitting for the free parameter $n_{\rm level}$, we essentially search for a correction to the adopted ``wrong'' PSD.} 

    \item Finally, a {\it third} case that we consider, is to use the Hyperbolic likelihood formulation ($\Lambda_{\cal H}$) of Eqs.~(\ref{eq:hypf_whitened},\ref{eq:ritheta}), again assuming the {\color{black} ``wrong" PSD} parameters of Eq. (\ref{eq:sim_psd_params_wrong}) for the noise.  {\color{black}Here, we allow the free parameters $\alpha$, $\delta$ of ($\Lambda_{\cal H}$) to offset the wrong assumption of the PSD. This application highlights one of the advantages of the Hyperbolic likelihood, which is the detection of any deviations from a given assumed noise model, or any kind of departures from the assumed Gaussianity of the residuals.
    }

\end{enumerate}

To validate the performance of each parameter estimation run, we compare them to the baseline case, which is the Gaussian likelihood with the correct noise model described by Eq.~(\ref{eq:sim_psd_model}) and the parameters of Eqs.~(\ref{eq:sim_psd_params}).

The results of this application are summarized in Figure~\ref{fig:ucbs_triangle_level} and in Table~\ref{tab:ucbs_level}. It is quite evident from the {\color{black} corner plot of the 2D posterior distributions in} Fig.~\ref{fig:ucbs_triangle_level}, that the naive approach 
{\color{black} 
of using the Gaussian likelihood $\Lambda_{\cal N}$ with a PSD based on the ``wrong" noise parameters of Eq. (\ref{eq:sim_psd_params_wrong}), yields {\rm underestimated posterior widths}.
On the other hand, when correcting the PSD, through the free $n_{\rm level}$ parameter of $\Lambda_{\cal W}$, we obtain results that agree with the baseline case. The $n_{\rm level}$ parameter is recovered as $1.5 \pm 0.12$ (90\% CI), which includes the correct value of $n_{\text {level }}^{\text {(correct) }} \sim 1.57$.
\begin{figure*}	\includegraphics[width=.9\linewidth]{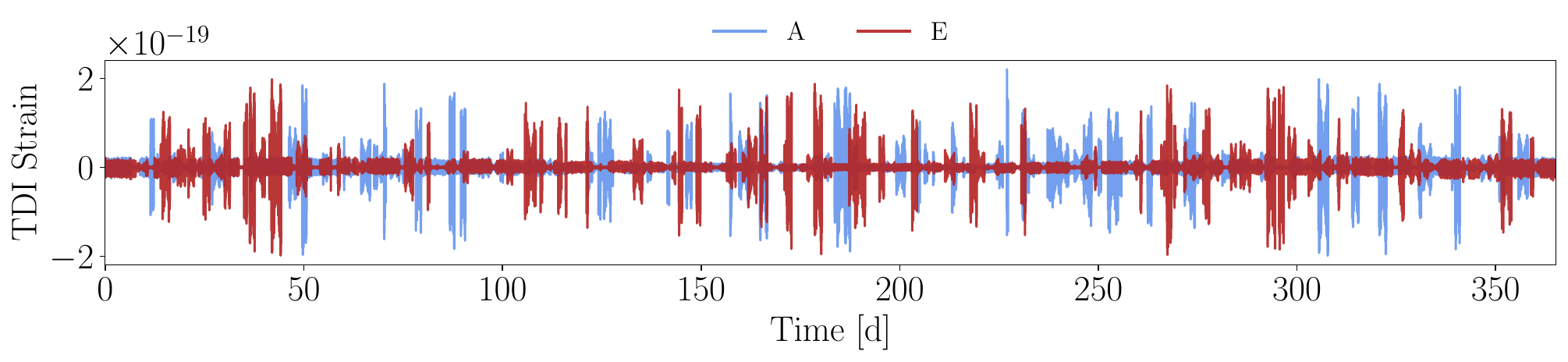}
 	\caption{ The time series of the $A$ (light blue) and $E$ (dark red) TDI channels include the injection binary signal, the nominal instrumental noise, and the noise bursts. The noise bursts are randomly placed in the synthetic time series, while their duration and spectral shapes are also generated by sampling from predefined distributions (see text for details). 
 \label{fig:ubcs_burst_data} }
\end{figure*}
Very similar 2D posterior distributions are also obtained with the Hyperbolic likelihood $\Lambda_{\cal H}$, but without assuming any parameterized model for the PSD. Instead, the correction of the wrong PSD levels is achieved through the free parameters $\alpha$, $\delta$ of $\Lambda_{\cal H}$. In Fig.~\ref{fig:ucbs_triangle_level} we also plot the $\log _{10} \delta-\log _{10} \alpha=\log _{10} n_{\text {level }}^{\text {(correct) }}$ line\footnote{\color{black} This equation is expected to hold, because the data are whitened with the PSD of the noise, see Eq. (\ref{eq:ritheta}).}, which aligns with the 2D posterior distributions of $\alpha$, $\delta$ as $\alpha \rightarrow \infty$ and $\delta \rightarrow \infty$, confirming that using $\Lambda_{\cal H}$ even with a wrong PSD in the inner product, one obtains posteriors equivalent to the baseline with the correct PSD.}

In Table~\ref{tab:ucbs_level} we present the Jensen Shannon divergence {\color{black}of the marginal posterior PDFs of all parameters, between the three different likelihood choices with wrong PSD levels and the baseline model of a Gaussian distribution with the correct PSD. It is quite evident, that the Hyperbolic distribution agrees really well with the baseline case and corrects the wrong PSD assumption with comparable accuracy as the Whittle likelihood.}

This result demonstrates the robustness of the Hyperbolic likelihood formulation in situations where the noise spectral shape is not entirely known, or when has features that are not modeled properly. As already discussed, such is the case of the future LISA data, where the confusion noise of the ensemble signal of all the UCBs in our Galaxy will generate a confusion stochastic foreground. Thus, a possible application would be to adopt a heavy-tailed likelihood, such as the one presented here, in order to perform parameter estimation and search under a robust statistical framework. 

As a final note, we should mention that we retrieved very similar results when inverting the noise assumptions for this exercise, i.e. the true noise being lower than the one assumed in the likelihood functions. In this case, the posterior spread was overestimated with the $\Lambda_\mrm{\cal N}$ likelihood and correctly estimated with the other two choices.

\begin{ruledtabular}
    \begin{table}[ht]
    \caption{Jensen-Shannon (JS) divergence {\color{black}of the marginal posterior PDFs of all parameters, between the three different likelihood choices with wrong PSD levels in Sec. \ref{sec:ucbs_level} and the baseline model of a Gaussian distribution with the correct PSD} (see text for details). \label{tab:ucbs_level}}
        \begin{tabular}{c*{4}{w{c}{2.44cm}}}
                            & & Jensen-Shannon Divergence $(\times10^{-3})$\\
                           $\param$ & $\Lambda_{\cal N}$ & $\Lambda_{\cal W}$ & $\Lambda_{\cal H}$ \\\hline 
            $\log_{10}\mathcal{A}$  & $11$ & $0.4$ & $0.3$ \\ 
            $f_0$                   & $12$ & $0.7$ & $0.4$ \\
            $\log_{10}(\dot{f}_0)$  & $1$ & $0.4$ & $0.9$ \\
            $\cos\iota$             & $11$ & $0.4$ & $0.4$ \\ 
            $\sin\beta_{\rm{E}}$             & $11$ & $0.4$ & $0.5$ \\
            $\lambda_{\rm{E}}$               & $7$ & $1.1$ & $1.7$ \\
            $\psi$                  & $12$ & $0.4$ & $0.6$ \\
            $\phi_0$                & $5$ & $0.9$ & $2.3$ \\
        \end{tabular} 
    \end{table}
\end{ruledtabular}

\subsubsection{Gaussian bursts \label{sec:ucbs_bursts}}

For our third application, we focus on a {\color{black}somewhat} more realistic scenario. In particular, we inject the data {\color{black} generated in Sec. \ref{sec:ucbs_level}} with bursts of Gaussian noise, placed randomly in the time series of the $A$ and $E$ TDI channels. The noisy bursts are generated by the same model of Eq.~(\ref{eq:sim_psd_model}) and with parameters drawn uniformly from {\color{black} $\log_{10}A\sim\mathcal{U} [-32, \, -28]$, and $\log_{10}P\sim\mathcal{U} [-23.6, \, -19.6]$ (in the units of Eq. (\ref{eq:sim_psd_params})). Their duration $d_\mrm{burst}$ is also sampled uniformly, as $d_\mrm{burst} \sim \mathcal{U}[ 0.01,\, 10 ]~\mrm{days}$. The resulting time series thus consists of the waveform model injected in Gaussian noise with the addition of the Gaussian bursts and is shown in Fig.~\ref{fig:ubcs_burst_data}}. Naturally, the presence of such noise non-stationarities is expected to distort the spectral shape of the overall noise PSD. However, in our application, we focus on a single UCB, which is very well localized in frequency. Thus the effect of the bursts is similar to {\color{black} the mismodeling of the PSD in} Section~\ref{sec:ucbs_level} above. This is illustrated in Fig.~\ref{fig:ubcs_burst_data_freq}, where {\color{black}the individual components of the signal, the simulated Gaussian instrumental noise, and the Gaussian noise bursts are shown in the frequency domain}.

\begin{figure*} 
\includegraphics[width=.9\linewidth]{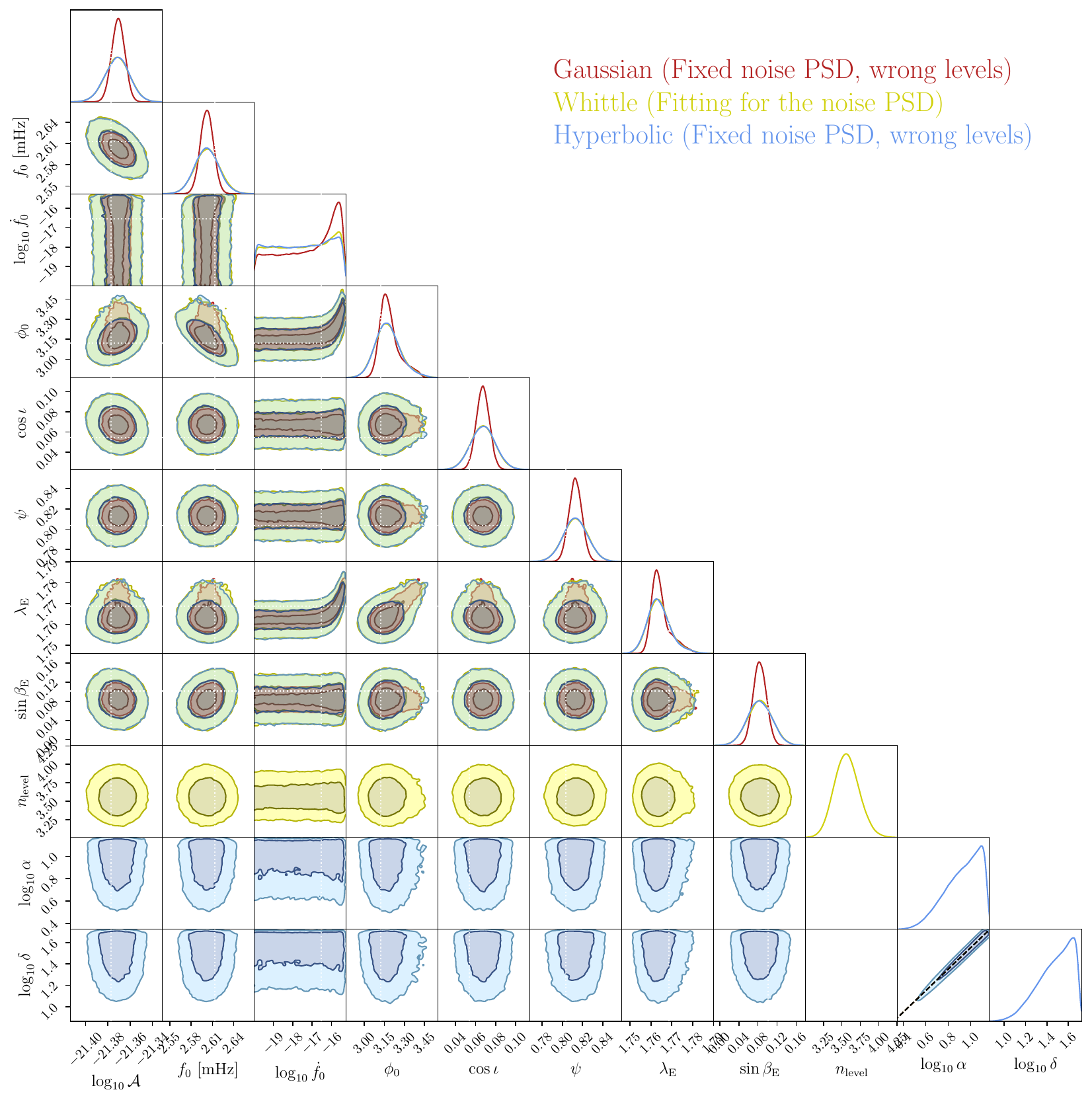}
 	\caption{Same as Fig. \ref{fig:ucbs_triangle_level}, but for the case of noise non-stationarities of Section~\ref{sec:ucbs_bursts}. {\color{black} The dashed black here represents the $n_\mrm{level}$ noise parameter as estimated with the Whittle likelihood analysis (see text for details).}}
\label{fig:ucbs_triangle_bursts}
\end{figure*}

Hence, we can adopt the same analysis methodology, which uses the same list of likelihood formulations that we used in previous sections. We first sample the posterior of the parameters given the data with a Gaussian likelihood, which assumes the artifacts-free noise model ($\Lambda_\mrm{\cal N}$). This assumption about the data is expected to yield a biased estimate of the waveform parameters posterior widths. We then repeat the analysis {\color{black} using the Whittle likelihood $\Lambda_\mrm{\cal W}$ and fitting for the noise level $n_\mrm{level}$}. Finally, we use the Hyperbolic distribution ($\Lambda_\mrm{\cal H}$), where we fit for the $\alpha$ and $\delta$ parameters of Eq.~(\ref{eq:hypf_whitened}). 
{\color{black}Notice that in this application}, we do not have a baseline run, because we do not have access to the ``true'' noise PSD {\color{black}due to the randomly injected Gaussian bursts}. 

The results can be summarized in Fig.~\ref{fig:ucbs_triangle_bursts}, where we plot the 2D posterior slices of the sampled parameters for each case. It is fairly obvious that the Hyperbolic and the noise-fitting Gaussian likelihoods present {\color{black}almost identical performance, with the corresponding marginal posterior distributions agreeing very well with each other}. This is verified when we compute the Jensen-Shannon divergence, which yields figures of order $\sim10^{-4}$ for all parameters. {\color{black} In addition, as expected, the $n_\mrm{level}$ noise parameter estimated with the Whittle likelihood (the maximum posterior value for $n_\mrm{level}$ is represented by the black dashed line in Fig.~\ref{fig:ucbs_triangle_bursts}), agrees with the $\delta/\alpha$ ratio recovered by the analysis using the hyperbolic likelihood. } This result confirms, once again, the versatility and robustness of the Hyperbolic likelihood $\Lambda_\mrm{\cal H}$ in parameter estimation situations, where the instrumental noise properties are not completely known.


\section{Conclusions and discussion\label{sec:conclusions}}

We have introduced a heavy-tailed likelihood framework for robust inference and demonstrated its applications in Gravitational Wave data analysis. In particular, we have adopted the Generalized Hyperbolic ($\mathcal{GH}$) distribution, which has a number of widely used distributions of the exponential family as limiting distributions. This built-in flexibility allows us to model the residual noise with more accuracy, which is  very useful when dealing with real data. Such cases are commonly encountered in GW Astronomy, where the GW signals need to be searched for in the noise of a given detector {\color{black}or of a network of detectors}. This task can be challenging due to the limited knowledge of the statistical properties of the detector noise, which might depart from the ideal Gaussian properties. This usually happens due to noise transients and bursts, data gaps, spectral lines, or even the presence of foreground stochastic GW signals. 

{\color{black}Based on a subclass of the ($\mathcal{GH}$) distribution (the Hyperbolic distribution, $\mathcal{H}$), we derived the Hyperbolic likelihood function $\Lambda_{\cal H}$ of Eq. (\ref{eq:hyp_general}), which depends on parameters $\alpha, \delta$ and $\bm{\beta}$} and demonstrated its performance and flexibility in simple test cases. We simulated random variables from distributions that are known to be limiting cases of $\mathcal{GH}$. Those are the Gaussian ($\mathcal{N}$), $\cal NIG$, and Student's t distributions. {\color{black}In each case, the inferred parameters $\alpha, \delta$ and $\bm{\beta}$ of $\Lambda_{\cal H}$ were demonstrated (as evidenced by the Jensen-Shannon divergence) to lead to a close match between the target and posterior distributions, even for cases where one of the $\mathcal{GH}$ parameters ($\lambda$) of the target distribution was different than the $\lambda$=1 value assumed in the Hyperbolic distribution. This is a nice demonstration that the Hyperbolic distribution has sufficient degrees of freedom to adjust (with acceptable accuracy) to different distributions.}

As as next step, we applied this methodology to more realistic applications, drawn from problems encountered in GW Astronomy. For the case of the current ground-based detectors, the spectral shape of the noise can be understood fairly well  to allow inference with relative confidence, due to the accessibility of the instrument and the numerous auxiliary channels available. Even in that case, there is always the risk of transient events, such as noisy glitches, which can happen randomly in time. The situation is quite different for future space-based observatories, such as LISA, where the spectral shape of the noise will not be completely known. This is due to the limited number of available auxiliary channels, as well as due to the ``confusion noise'' generated by the ensemble of the vast number of signals detectable with LISA. 

For demonstration purposes, we limited our analyses to a single signal from a Double White Dwarf binary, {\color{black}injected in colored Gaussian noise representative of the LISA detector} with relatively high SNR ($\rho = 122.7$, see Table~\ref{tab:ucb_params} for its waveform parameters) and investigated three different applications. {\color{black}In the first investigation, 
we performed
parameter estimation using the Hyperbolic likelihood $\Lambda_{\cal H}$ we propose in Eqs.~(\ref{eq:hypf_whitened}),  (\ref{eq:ri_nonwhite}), where the level of the noise was assumed completely unknown (no PSD was assumed and no whitening was performed in the frequency domain). From the inferred parameters $\{\alpha, \,\delta\}$ of the symmetric Hyperbolic distribution, we were able to characterize the noise.} In particular, as theory predicts, we found that the distribution of the residuals asymptotically (as $\alpha,\delta\rightarrow\infty$) tends to the assumed Gaussian distribution, with $\log\delta - \log\alpha = \log S_n^{\mrm{A}\mrm{E}}(f_0)$, where $S_n^{\mrm{A}\mrm{E}}(f_0)$ is the spectral noise density at the emission frequency $f_0$  of the binary. 

Then, we investigated a case of mismodeling the PSD levels of the noise. Essentially, we assume a wrong noise level for the parameter estimation process, which is then plugged into the Gaussian likelihood of Eq.~(\ref{eq:gaussian_llh}). For this investigation, we assumed that the noise PSD was lower than the true value that was used to simulate the data. As before, we also performed the parameter estimation analysis of a single Galactic Binary using the Hyperbolic likelihood $\Lambda_\mrm{\cal H}$, with the expectation that the $\{\alpha, \,\delta\}$ coefficients would ``tune'' the likelihood to mitigate for this mismodeling of the noise PSD. This was verified with the results presented in Fig.~\ref{fig:ucbs_noise_fit} and Table~\ref{tab:ucbs_level}. The estimated waveform posteriors for the parameters, when using $\Lambda_\mrm{\cal H}$, are almost identical to those obtained with the baseline Gaussian likelihood using the correct noise levels. In addition, the ratio $\delta/\alpha$ obtained from posteriors of the parameters of the hyperbolic likelihood converged to the ratio $n_\mrm{level}^{(\mrm{true})}$ between the true noise PSD level and the mismodeled case. {\color{black}The true PSD was thus fully characterized using the $\Lambda_\mrm{\cal H}$ likelihood.}

Finally, we tested this framework in a somewhat more realistic scenario, where the instrumental noise was ``polluted'' with Gaussian bursts, randomly placed in time for both TDI channels considered in the analysis. Using the same approach as before, we obtained results that verify the functionality of the heavy-tailed framework introduced in this work. In particular, we demonstrated that with the hyperbolic likelihood, we were able to recover the correct posterior widths for the waveform parameters, {\color{black}which were larger than predicted by the simple Gaussian likelihood,} due to the additional uncertainty induced by the noisy bursts (see Section~\ref{sec:ucbs_bursts} for details).

We expect that the new framework we introduced in this work will be very useful for data from future observatories, where the noise model will not be completely accessible. The Hyperbolic likelihood is quite versatile, as {\color{black}the shape of the Hyperbolic distribution can  adjust to fit the residual data. In that sense, an advantage to the Whittle approximation of the likelihood, is that the Hyperbolic likelihood can converge to a different distribution than the Gaussian}. Thus, there are many applications where this framework can be utilized, a primary example being the modeling of noise artifacts, such as bursts, data gaps, glitches, or other types of non-stationarities. At the same time, even in the absence of noise non-stationarities, the Hyperbolic likelihood can be employed in order to probe any mismodeling of our noise models. Finally, another application would be to construct Gaussianity tests on different segments of data, based on the values of the parameters of the Hyperbolic likelihood and tools such as the shape triangle. 

In our computations, we used Parallel Tempering MCMC methods~\cite{eryn}, and GPU accelerated waveforms~\cite{katzucbszenodo} and we make our codes available as open source software\footnote{\url{https://github.com/asasli/hyperbolic_likelihood_filter}}.


\begin{acknowledgments}

We wish to thank the entire LDC Group, and especially M. Le Jeune, for their useful comments and very helpful guidance with the LDC data. We also thank M. Katz, Q. Baghi, and N. Korsakova for their useful comments and fruitful discussions. NS and NK acknowledge  support from the Gr-PRODEX 2019 funding program (PEA 4000132310). NK acknowledges the funding from the European Union’s Horizon 2020
research and innovation programme under the Marie Skłodowska-Curie grant agreement No 101065596. AS acknowledges the Bodossaki Foundation for support in the form of a PhD scholarship.

\end{acknowledgments}

\appendix
{\color{black}\section{The shape triangle \label{sec:shapetriangle}}

The shape triangle is a very useful graphical tool for visualizing the degree to which a distribution belonging to the exponential family is heavy-tailed or skewed. One can begin by using the Generalized Hyperbolic distribution to characterize a new distribution. Then, based on the recovered $\{\alpha,\,{\bm \beta},\,\delta\}$ coefficients, one can qualitatively categorize the distribution of interest by placing it into a shape triangle~\cite{Prause1999TheGH,KUCHLER19991}. To do that, we use a different parametrization from the usual $\{\alpha,\,{\bm \beta},\,\delta\}$ set. Following~\cite{Prause1999TheGH,KUCHLER19991}, we can compute
\begin{equation}
\begin{split}
\zeta = \delta\sqrt{\alpha^2-\beta^2},  \quad \varrho = \beta / \alpha,  \\
\xi = (1 + \zeta)^{-1/2}, \quad \chi = \xi\varrho, \\
\label{eq:triangle_params}
\end{split} 
\end{equation}
and then use the scale and location-invariant parameters $\chi$ and $\xi$, which form a triangle in the $\chi-\xi$ plane, since $0 \leq \abs{\chi} < \xi < 1 $. The position inside the shape triangle gives us visual information about the distribution of heavy-tailedness and skewness. This visual tool can be helpful in practical applications, such as monitoring data quality in real time or classifying different data segments, depending on their statistical properties~\cite{sed}.

Here, we investigate the same test distributions as in Sec.~\ref{sec:apps_and_usage}, but for the two cases, for which the parameter $\lambda$ is negative, we adopt the $\cal GH$ distribution for $\lambda<0$ as in~\cite{Prause1999TheGH}
\begin{equation}
\begin{split} 
    \Lambda_{\cal GH_{\rm d}}(\alpha, \delta) = &n\Big[ - \lambda\log(\delta) + \frac{d}{2}{\rm log} (2\pi)  - \log \left(K_{\lambda} (\delta\alpha)\right)\Big] \\
    &- \frac{1}{2}\left(\lambda - \frac{d}{2}\right)\sum^n_{j=1}(\delta^2 + x_j^2) \\
    &+ \sum^n_{j=1}K_{\lambda-\frac{d}{2}} \left(\alpha\sqrt{\delta^2 + x_j^2}\right).
\label{eq:hyp_lambda_negative}
\end{split} 
\end{equation}
The reason that we assume two different likelihoods, one of the positive values of $\lambda$ ($\Lambda_{\cal H}$ of Eq.~(\ref{eq:hyp_general})) and another one for the negative values ($\Lambda_{\cal GH_{\rm d}}$ of Eq.~(\ref{eq:hyp_lambda_negative})), arises from the definition of the domain in~Eq. (\ref{eq:domain}). Thus, the difference here compared to Sec.~\ref{sec:apps_and_usage} is that the posteriors for the $\{\alpha,\,\beta,\,\delta\}$ are estimated using the $\Lambda_{\cal GH_{\rm d}}$ Eq.~(\ref{eq:hyp_lambda_negative}) when the test-case distribution is either the Student's t-distribution or $\cal NIG$.}

After we obtain the posteriors, we calculate the $\{\chi, \xi\}$ parameters from Eqs.~(\ref{eq:triangle_params}) and place them onto the shape triangle, which is shown in Fig.~\ref{fig:toy_triangle}. The shape triangle, gives us a graphical representation (on the $\chi-\xi$ plane) of the characteristics of the resulting distribution. We have marked the injected values with crosses, while the recovered parameters are marked with colored dot markers ($\mathcal{N}$ with yellow, $\mathcal{H}$ with purple, ${\cal NIG}$ with red and, and finally the Student's t-distribution case with light blue). From the results displayed in Figure~\ref{fig:toy_triangle}, it is evident that we were able to recover the underlying distributions as special or limiting cases of the $\cal GH$ distribution. 

\begin{figure} [ht]
     \centering
         \includegraphics[width=0.45
\textwidth]{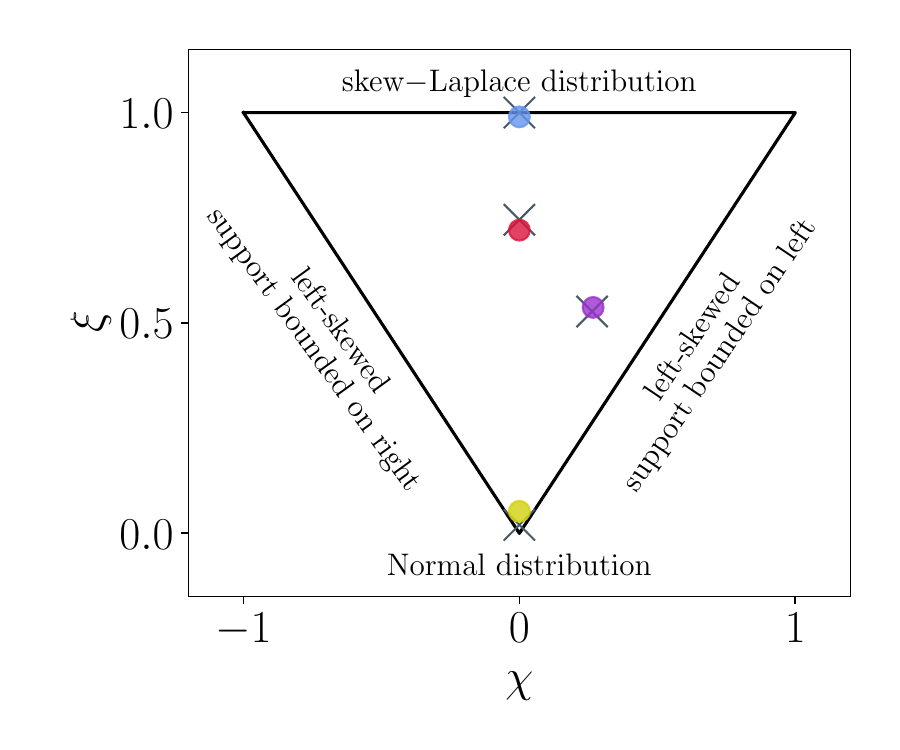}
        \caption{ Triangle shape for the injected values of $\chi, \xi$ ( crosses) and the recovered values (filled circles) for the data generated assuming Gaussian $\cal N$ (yellow), Hyperbolic $\cal H$ (purple), Normal Inverse Gaussian $\cal NIG$ (red) and Student's t (light blue) distributions.}
                 \label{fig:toy_triangle}
\end{figure}
\providecommand{\noopsort}[1]{}\providecommand{\singleletter}[1]{#1}%

\end{document}